\documentclass[dvipdfmx]{article}
\usepackage{amsmath,amsfonts,amssymb}
\usepackage{graphicx}

\usepackage{braket}
\newcommand{\tr}{{}^t\!}
\newcommand{\oC}{\mathcal{C}}
\newcommand{\oD}{\mathcal{T}}
\newcommand{\oH}{\mathcal{H}}
\newcommand{\oM}{\mathcal{M}}
\newcommand{\oO}{\mathcal{O}}
\newcommand{\op}{\mathcal{D}}
\newcommand{\oP}{\mathcal{P}}
\newcommand{\oR}{\mathcal{R}}
\newcommand{\oT}{\Theta}
\newcommand{\oU}{\mathcal{U}}
\newcommand{\lleft}{\left.\left}
\newcommand{\rright}{\!\!\!\right)\!\!\!\!\right}

\begin{document}

\title{Decomposition of the Hessian matrix for action\\
at choreographic three-body solutions\\
with figure-eight symmetry
}
\author{Toshiaki Fujiwara, Hiroshi Fukuda\\
{\it College of Liberal Arts and Sciences, Kitasato University}\\[3mm]
Hiroshi Ozaki\\
{\it Laboratory of general education for science and technology,}\\
{\it Faculty of Science, Tokai University}\\[3mm]
e-mail: 
fujiwara@kitasato-u.ac.jp,
fukuda@kitasato-u.ac.jp\\
and ozaki@tokai-u.jp
}
\date{}
%
%

\maketitle

\begin{abstract}
We developed a method to calculate the eigenvalues and eigenfunctions of the second derivative (Hessian) of action at choreographic three-body solutions 
that have the same symmetries as the figure-eight solution.  
A choreographic three-body solution is a periodic solution to equal mass  planar three-body problem under potential function 
$\sum_{i<j} U(r_{ij})$, in which three masses chase each other on a single closed loop
with equal time delay.

We treat choreographic solutions
that have the same symmetries as the figure-eight,
namely, symmetry for choreography, for time reversal, and for time shift of half period.
The function space  of periodic functions
are decomposed into  
five  subspaces by these symmetries.
Namely, 
one subspace of trivial oscillators with eigenvalue $4\pi^2/T^2\times k^2$, $k=0,1,2,\dots$,
four subspaces of choreographic functions,
and four subspaces of ``zero-choreographic'' functions.
Therefore, the matrix representation of the Hessian is also decomposed into 
nine corresponding  blocks.
Explicit expressions of base functions
and the matrix representation of the Hessian
for each subspaces are given.

The trivial eigenvalues with $k\ne 0$ are quadruply degenerated,
while with $k=0$
are doubly degenerated
that correspond to the conservation of linear momentum in $x$ and $y$ direction.
The  eigenvalues in choreographic subspace have no degeneracy in general.
In  ``zero-choreographic'' subspace,
every eigenvalues  are doubly degenerated.
\end{abstract}

\newpage
\tableofcontents

\newpage
\section{Definitions and notations}
Consider a planar three-body problem with equal masses $m_\ell=1$,
$\ell=0,1,2$,
defined by the  Lagrangian
\begin{equation}
\label{Lagrangian}
L=\frac{1}{2}\sum_\ell \left(\frac{dq_\ell}{dt}\right)^2 + \sum_{i \ne j} V(|q_i-q_j|),\ 
\end{equation}
$q_\ell=(q_\ell^x, q_\ell^y) \in \mathbb{R}^2$.
The potential $V$ may be an extended Newton potential
\begin{equation}
\label{homogeneousPotential}
V_\alpha(|q_i-q_j|)=
	\begin{cases}
		\alpha^{-1}/|q_i-q_j|^\alpha	& (\alpha \ne 0)\\
		\log |q_i-q_j|				& (\alpha=0)
	\end{cases}
\end{equation}
or a Lennard-Jones type potential
\begin{equation}
V=\frac{a}{|q_i-q_j|^\alpha} -\frac{b}{|q_i-q_j|^\beta}.
\end{equation}

Choreographic solutions are periodic solutions to the equation of motion
\begin{equation}
\frac{d^2 q_\ell}{dt^2} = \frac{\partial}{\partial q_\ell}\sum_j V(|q_j-q_\ell|)
\end{equation}
that satisfy
\begin{equation}
(q_0,q_1,q_2)=(q(t), q(t+T/3), q(t-T/3)),
\end{equation}
where $T$ is the period.

Some choreographic solutions are known.
The famous figure-eight solution was discovered numerically by 
C.~Moore \cite{Moore}  in 1993.
Then, A.~Chenciner and R.~Montgomery \cite{CM} in 2000
rediscovered this  solution and gave a proof for existence.
M.~\v{S}uvakov and V.~Dmitra\v{s}inovi\'{c} in 2013
found many ``slalom solutions''.
L.~Sbano \cite{Sbano} in 2004 found the figure-eight solution
under the Lennard-Jones type potential.
H.~Fukuda, T.~Fujiwara and H.~Ozaki  \cite{LJ1} in 2017 found
many choreographic solutions
under the Lennard-Jones type potential.
Almost all these choreographic solutions have the same symmetries
as the figure-eight solutions.
Exceptional cases are $k=$ even slaloms,
where $k$ is the ``power of slalom'' \cite{SuvakovShibayama}.
The symmetries that we use in this note
will be shown in the section \ref{sectionSymmetries}.
Method in this note is applicable to 
the original figure-eight solution, $k=$ odd slaloms,
figure-eight solutions by Sbano et al., and Fukuda et al.


\subsection{Definition of the Hessian}
Consider the variation of action integral at a solution $q_\ell(t)$ to second order of 
variation $q_\ell(t) \to q_\ell(t)+\delta q_\ell(t)$,
\begin{equation}
S[q+\delta q]=S[q] + \frac{1}{2}\int_0^T dt \Bigg(
	\sum_{\ell}\left(\frac{ d\delta q_\ell}{dt}\right)^2 
	+ \sum_{i,j}\delta q_i  \left(\frac{\partial^2}{\partial q_i \partial q_j}\sum V\right) \delta q_j
	\Bigg).
\end{equation}
Here,
the first order term $\delta S$ is zero as $q_\ell$ is a solution.
We restrict ourselves to consider the variational function $\delta q_\ell$ be 
periodic function with period $T$. Then, partial integration makes
\begin{equation}
\int_0^T dt \sum_\ell \left(\frac{ d\delta q_\ell}{dt}\right)^2 
=\int_0^T dt \sum_\ell \delta q_\ell \left(-\frac{d^2}{dt^2}\right)\delta q_\ell.
\end{equation}
Thus, the variation is given by
\begin{equation}
S[q+\delta q]=S[q] + \frac{1}{2}\int_0^T dt\ {}^t\! \Psi \oH \Psi,
\end{equation}
where
\begin{equation}
\Psi
=\left(\begin{array}{c}\delta q_0^x \\\delta q_0^y \\
	\delta q_1^x \\\delta q_1^y \\
	\delta q_2^x \\\delta q_2^y\end{array}\right)
\end{equation}
is $\mathbb{R}^6$ column vector
and $\oH$ is the Hessian that has the form
\begin{equation}
\begin{split}
\label{defOfHessian}
\oH= -\frac{d^2}{dt^2} + {}^t\!\Delta \oU \Delta,\ 
\oU=\left(\begin{array}{ccc}u_{12} & 0 & 0 \\0 & u_{20} & 0 \\0 & 0 & u_{01}\end{array}\right).
\end{split}
\end{equation}
Here, $u_{ij}$ is $2\times 2$ matrix and $\Delta$ is $6\times 6$ anti-symmetric matrix
\begin{equation}
u_{ij}=\left(\begin{array}{ll}
		u_{ij}^{xx} & u_{ij}^{xy} \\
		u_{ij}^{yx} & u_{ij}^{yy}
	\end{array}\right),
\Delta = \left(\begin{array}{rrr}
0 & E_2 & -E_2 \\-E_2 & 0 & E_2 \\E_2 & -E_2 & 0\end{array}\right),
E_2=\left(\begin{array}{cc}1 & 0 \\0 & 1\end{array}\right),
\end{equation}
and ${}^t\!\Delta$ stands for the transpose of $\Delta$.
The explicit expression of $u_{ij}$ for extended Newton potential, for example, is
\begin{equation}
\begin{split}
u_{ij}&=\frac{\alpha+2}{r_{ij}^{\alpha+4}}
\left(\begin{array}{cc}(q_i^x-q_j^x)^2 & (q_i^x-q_j^x)(q_i^y-q_j^y)\\
	(q_i^x-q_j^x)(q_i^y-q_j^y) & (q_i^y-q_j^y)^2\end{array}
	\right)
	-\frac{1}{r_{ij}^{\alpha+2}}E_2,\\
r_{ij}&=|q_i-q_j|.
\end{split}
\end{equation}
When there is no confusion, we will use the following abbreviated notations
\begin{equation}
\Psi
=\left(\begin{array}{c}\delta q_0 \\\delta q_1 \\\delta q_2\end{array}\right)
\mbox{ and }
\Delta = \left(\begin{array}{rrr}0 & 1 & -1 \\-1 & 0 & 1 \\1 & -1 & 0\end{array}\right),
\end{equation}
and similar.
In this note, we consider the eigenvalue problem
\begin{equation}
\label{eigenValueProblem}
\oH \Psi = \lambda \Psi.
\end{equation}

For linear operator $\mathcal{O}$, we express its eigenvalue $\mathcal{O}'$.
If $\mathcal{O}$ satisfies $\mathcal{O}^2=1$, the eigenvalue is one of 
$\mathcal{O}'=\pm 1$.
While,
if $\mathcal{O}^3=1$ is satisfied, then $\mathcal{O}'=1, \omega, \omega^2$
with $\omega = (-1\pm i\sqrt{3})/2$.

For any two functions $\Phi$, $\Psi$ and any operator $\mathcal{O}$,
we define and write the inner product
\begin{equation}
\label{defOfInnerProduct}
\braket{\Phi|\Psi}=\frac{2}{3T}\int_0^T dt\  \tr\Phi(t)\Psi(t),\ 
\braket{\Phi|\mathcal{O}|\Psi}=\frac{2}{3T}\int_0^T dt\  \tr\Phi(t)\mathcal{O}\Psi(t).
\end{equation}
The factor $2/(3T)$ is for later convenience.
The inner product satisfy
\begin{equation}
\braket{\Phi|\Psi}=\braket{\Psi|\Phi},\ 
\braket{\Phi|\oO|\Psi}
=\braket{\Phi|\oO\Psi}
=\braket{\tr\oO\Phi|\Psi}.
\end{equation}

For writing basis vectors,
it is convenient to write $x$ and $y$ basis together
in a matrix form,
\begin{equation}
\left(\begin{array}{cc}
x_1 & 0 \\
0 & y_1 \\
x_2 & 0\\
0 & y_2 \\
x_3 & 0 \\
0 & y_3\\
\end{array}\right),
\end{equation}
with
\begin{equation}
\frac{2}{3T}\int_0^T dt (x_1^2+x_2^2+x_3^2)
=\frac{2}{3T}\int_0^T dt (y_1^2+y_2^2+y_3^2)
=1.
\end{equation}
Then
\begin{equation}
a^x \left(\begin{array}{c}
x_1\\
0\\
x_2\\
0\\
x_3\\
0\\
\end{array}\right)
+a^y \left(\begin{array}{c}
0\\
y_1\\
0\\
y_2\\
0\\
y_3\\
\end{array}\right)
=\left(\begin{array}{cc}
x_1 & 0 \\
0 & y_1 \\
x_2 & 0\\
0 & y_2 \\
x_3 & 0 \\
0 & y_3\\
\end{array}\right)
\left(\begin{array}{c}
a^x\\
a^y
\end{array}\right).
\end{equation}
If $x_k=y_k=z_k$, we simply write
\begin{equation}
\left(\begin{array}{cc}
z_1 & 0 \\
0 & z_1 \\
z_2 & 0 \\
0 & z_2 \\
z_3 & 0 \\
0 & z_3
\end{array}\right)
=
\lleft(\begin{array}{c}
z_1 \\z_2 \\z_3\end{array}\rright).
\end{equation}
It should be stressed that
this is not a constraint but a matrix representation of two independent basis vectors.
For example,
\begin{equation}
\Phi_n
=
\lleft(\begin{array}{l}
\cos(n\nu t) \\\cos(n\nu t + 2\pi/3) \\\cos(n\nu t - 2\pi/3)\end{array}\rright)
\end{equation}
is a matrix representation of two independent basis
\begin{equation}
\left(\begin{array}{l}
\cos(n\nu t)\\
0\\
\cos(n\nu t + 2\pi/3)\\
0\\
\cos(n\nu t - 2\pi/3)\\
0
\end{array}\right)
\mbox{ and }
\left(\begin{array}{ll}
0\\
\cos(n\nu t)\\
0\\
\cos(n\nu t + 2\pi/3)\\
0\\
\cos(n\nu t - 2\pi/3)
\end{array}\right)
\end{equation}
and $\braket{\Phi_m|\Phi_n}=\delta_{mn}$
for $\nu=2\pi/T$.

\subsection{Decomposition of function space by orthogonal projection operator}
A linear operator $\oP$ is called ``orthogonal projection operator''
if it  satisfies
\begin{equation}
\oP(\oP-1)=0 \mbox{ and } \tr\oP=\oP.
\end{equation}
This operator decompose function space $W$ into two orthogonal subspaces
$W_1$ and $W_0$,
\begin{equation}
W_1 = \{ f |f \in W, \oP f = f\}, W_0=\{ f | f \in W, \oP f =0\},
W=W_1 \oplus W_0.
\end{equation}
\begin{figure}
   \centering
   \includegraphics[width=6cm]{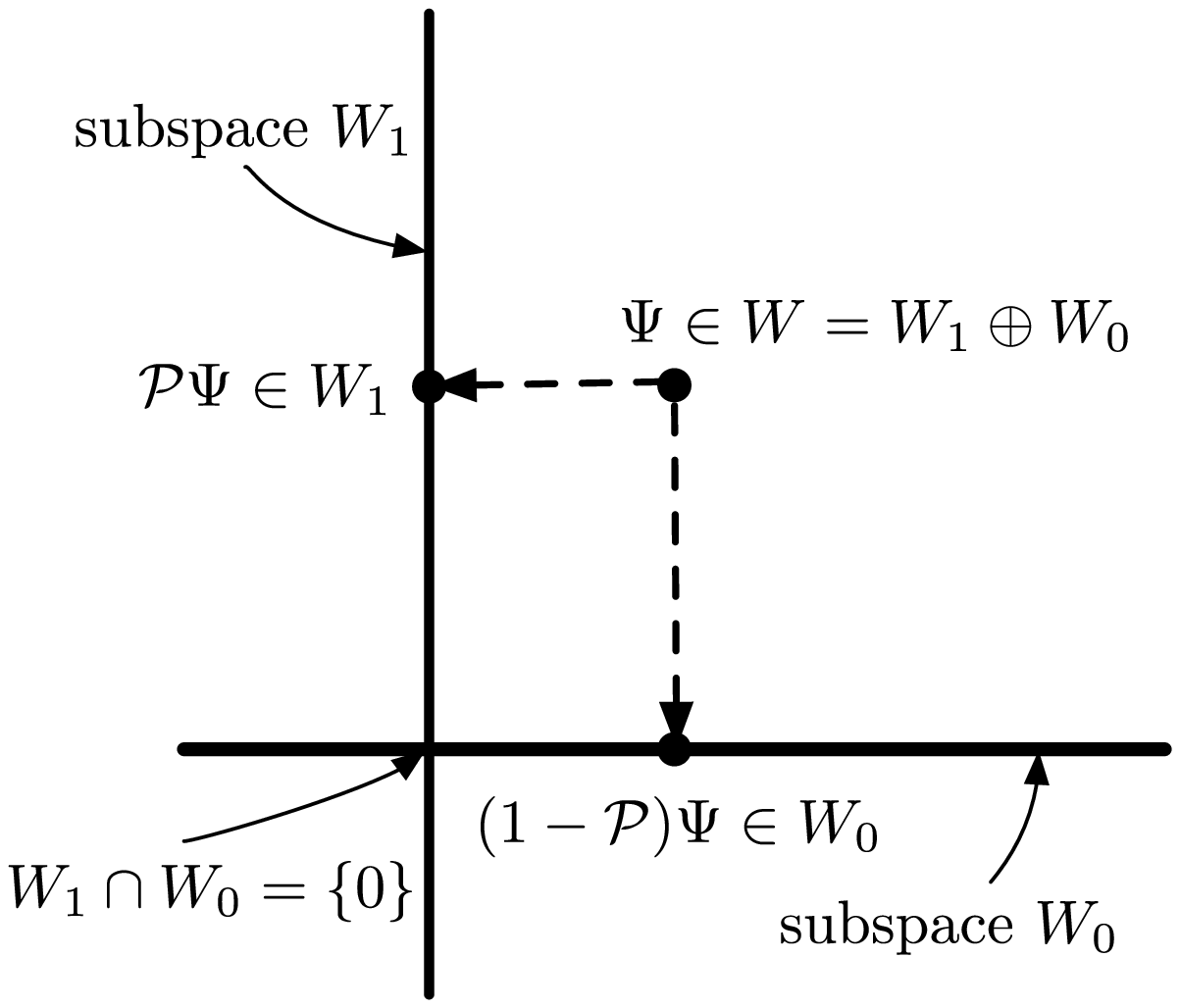}
   \caption{An schematic image of decomposition of function space
   $W=W_1 \oplus W_0$
   by orthogonal projection operator $\oP$ that satisfy
   $\oP(\oP-1)=0$, $\tr \oP=\oP$}
   \label{fig:decompositionOfFunctionSpace}
\end{figure}
The subspaces $W_1$ and $W_0$ are mutually orthogonal compliment,
because arbitrary elements $f \in W_1$ and $g \in W_0$  are orthogonal
by $\braket{g|f}=\braket{g|\oP f}=\braket{\tr\oP g|f}=0$,
and arbitrary element $\Psi \in W$ will be decomposed by the identities
$\Psi = \oP \Psi + (1-\oP) \Psi$.
Here,
$\oP \Psi \in W_1$ and $(1-\oP) \Psi \in W_0$ since
$\oP(\oP \Psi )=\oP \Psi$ and $\oP((1-\oP) \Psi )=0$.
See figure \ref{fig:decompositionOfFunctionSpace}.

If the Hessian $\oH$ is commutable with $\oP$,
matrix representation of the Hessian is also decomposed into two blocks
for $W_1$ subspace and $W_0$ subspace,
because
the matrix elements of $\oH$ between $W_1$ subspace and $W_0$ subspace are zero,
\begin{equation}
\braket{g|\oH|f}=\braket{g|\oH|\oP f}=\braket{\tr \oP g|\oH|f}=0
\mbox{ for } f \in W_1 \mbox{ and } g \in W_0.
\end{equation}

In the following sections,
we will find  some orthogonal projection operators
that are commutable each other and $\oH$.
These projection operators decompose the function space into small subspaces
and
make the matrix representation of  $\oH$ small blocks.

\section{Four zero eigenvalues and  quadruply degenerated trivial eigenvalues}
\label{section:zeroAndTrivialMode}
In this section,
we describe $4$ zero eigenvalues and  quadruply degenerated trivial eigenvalues,
which  always exist for the Hessian (\ref{defOfHessian}).

Corresponding to the conservation law,
the linear momentum for $x$ and $y$ direction, angular momentum
and the energy,
the Hessian always has $4$ zero eigenvalues
for variations
$\delta q_\ell = \tr(1,0)$, $\tr(0,1)$, $\tr(-q_\ell^y, q_\ell^x)$ and $ dq_\ell/dt$.

Since the function space is composed by all periodic functions with period $T$,
it contains a  trivial eigenfunction $\Psi(t) = {}^t\!(\delta q(t),\delta q(t),\delta q(t))$
that describes three bodies move coherently.
Since, $\Delta \Psi=0$ for this function,
the eigenvalue problem is reduced to the simple form
\begin{equation}
H \Psi
= -\frac{d^2}{dt^2} \Psi
= \lambda \Psi,
\end{equation}
that yields quadruply degenerated eigenvalue $\lambda = 4\pi^2/T^2 \times k^2$.
Namely
$\delta q=\tr(\cos 2\pi k t/T, 0)$, $\tr(0, \cos 2\pi k t/T)$,
$\tr(\sin 2\pi k t/T, 0)$, $\tr(0, \sin 2\pi k t/T)$
for positive integers $k$.
The two trivial functions for $k=0$ are already counted 
for translational function in the previous paragraph.

Let $\sigma$ be the cyclic permutation operator
$\sigma\ {}^t\!(\delta q_0, \delta q_1, \delta q_2)={}^t\!(\delta q_1, \delta q_2, \delta q_0)$,
or in the matrix form,
\begin{equation}
\sigma=\left(\begin{array}{ccc}
	0 & 1 & 0 \\0 & 0 & 1 \\1 & 0 & 0\end{array}\right)
\mbox{ and } {}^t\!\sigma = \sigma^{-1}.
\end{equation}
Since $\sigma^3=1$,
it defines the orthogonal projection operator
\begin{equation}
\oP_\sigma=\frac{1}{3}(1+\sigma+\sigma^2),\ 
\oP_\sigma(\oP_\sigma-1)=0.
\end{equation}
This operator splits the function space into
subspace with the eigenvalue
$\oP_\sigma'=1$ or $\oP_\sigma'=0$.
The subspace whose element has $\oP_\sigma'=1$ is the trivial subspace.
The complemental subspace is composed of functions
 with $\oP_\sigma'=0$.
 We call this subspace ``zero-centre-of-mass'' subspace,
 because each function in this subspace keeps the center of mass at the origin,
 \begin{equation}
 0
 =\frac{1}{3}(1+\sigma+\sigma^2)
 	\left(\begin{array}{c}\delta q_0 \\\delta q_1  \\\delta q_2 \end{array}\right)
=\frac{1}{3}
	\left(\begin{array}{c}
		\delta q_0+\delta q_1+\delta q_2\\
		\delta q_0+\delta q_1+\delta q_2\\
		\delta q_0+\delta q_1+\delta q_2
	\end{array}\right).
 \end{equation}
The figure-eight solution belongs to zero-center-of-mass subspace.

Now, let us proceed to the symmetry of the Hessian
at choreographic solutions
with figure-eight symmetry.

\section{Symmetries of the figure-eight solution}
\label{sectionSymmetries}
The figure-eight solution with proper $x, y$ coordinates and origin of time
has the following symmetries,
\begin{equation}
\label{choreographicSymmetry}
\mbox{choreographic symmetry: }
\left(\begin{array}{c}
q_0(t+T/3) \\q_1(t+T/3) \\q_2(t+T/3)
\end{array}\right)
=
\left(\begin{array}{c}
q_1(t) \\q_2(t) \\q_0(t)
\end{array}\right),
\end{equation}
\begin{equation}
\label{timeReversalSymmetry}
\mbox{time reversal symmetry: }
\left(\begin{array}{c}
q_0(-t) \\q_1(-t) \\q_2(-t)
\end{array}\right)
=
-\left(\begin{array}{c}
q_0(t) \\q_2(t) \\q_1(t)
\end{array}\right),
\end{equation}
and
\begin{equation}
\label{mirrorSymmetry}
\mbox{time shift in $T/2$:}
\left(\begin{array}{c}
q_0(t+T/2) \\q_1(t+T/2) \\q_2(t+T/2)
\end{array}\right)
=
-\left(\begin{array}{c}
\mu\ q_0(t) \\\mu\ q_1(t) \\ \mu\ q_2(t)
\end{array}\right),
\mu \left(\begin{array}{c}x \\y\end{array}\right)
=\left(\begin{array}{c}-x \\y\end{array}\right).
\end{equation}


\subsection{Choreographic symmetry}
Let $\oR^{1/3}$ be a time displacement operator
$\oR^{1/3}  \delta q(t) =\delta q(t+T/3)$
and $\oC=\sigma^{-1}\oR^{1/3}$.
Then the choreographic symmetry in (\ref{choreographicSymmetry}) is equivalent to
\begin{equation}
\oC
\left(\begin{array}{c}
q_0(t) \\q_1(t) \\q_2(t)
\end{array}\right)
=\left(\begin{array}{c}
q_0(t) \\q_1(t) \\q_2(t)
\end{array}\right).
\end{equation}
\begin{figure}
   \centering
   \includegraphics[width=8cm]{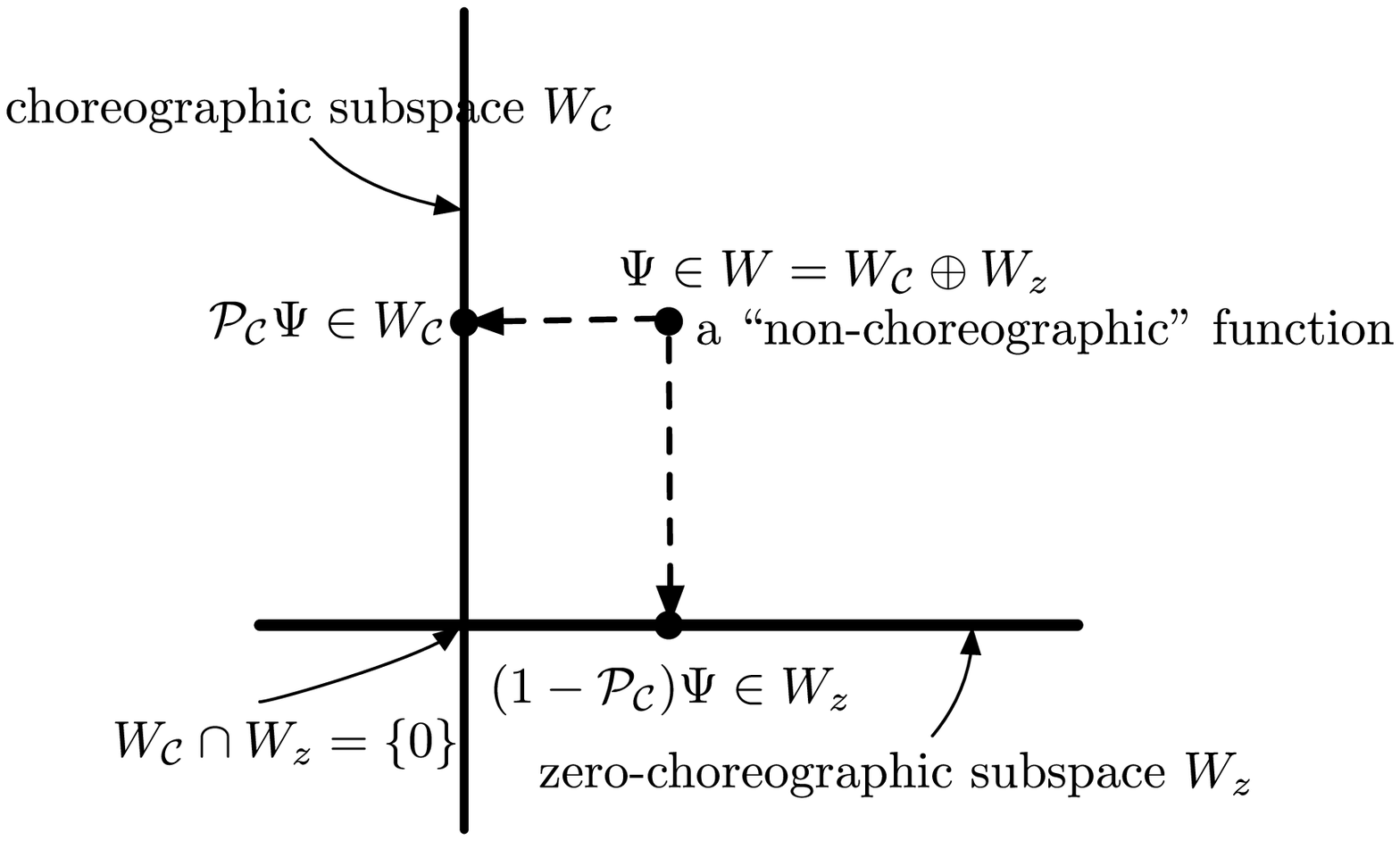}
   \caption{Decomposition of function space $W=W_\oC \oplus W_z$
   by orthogonal projection operator $\oP_\oC$.
  We call a function $\Psi$ ``non-choreographic'' \emph{function} if $(1-\oP_\oC)\Psi \ne 0$.
  While, we call the subspace $W_z$ ``zero-choreographic'' \emph{subspace}.
   }
   \label{fig:decompositionOfFunctionSpaceByOPC}
\end{figure}

Since the function space has period $T$,
\begin{equation}
\int_0^T dt f(t)g(t+T/3)=\int_0^T dt f(t-T/3)g(t).
\end{equation}
Namely, $\tr R^{1/3}=R^{-1/3}$
and $\tr\oC=\oC^{-1}$.
Since the operator $\oC$ satisfies $\oC^3=1$,
it defines the orthogonal projection operator
\begin{equation}
\oP_\oC=\frac{1}{3}(1+\oC+\oC^2),\ 
\oP_\oC(\oP_\oC-1)=0.
\end{equation}

Thus the operator $\oP_\oC$ decomposes
the function space $W$
into two orthogonal subspaces;
the choreographic subspace $W_c$
that made of functions that belong to the eigenvalue $\oP_\oC'=1$
(i.e. $\oC'=1$) and its orthogonal complement.
We call this complement  ``zero-choreographic'' subspace $W_z$.
\begin{equation}
\begin{split}
&W_c=\set{\Psi | \oP_\oC \Psi = \Psi},\ 
W_z=\set{\Psi | \oP_\oC \Psi = 0},\\
&W=W_c \oplus W_z 
\end{split}
\end{equation}
The figure-eight solution belongs to choreographic subspace.
\emph{In this note, we call the subspace $W_z$ ``zero-choreographic'' subspace
to avoid  confusion with ``non-choreographic'' function.}
A function is called ``non-choreographic'',
if it is not choreographic.
See figure \ref{fig:decompositionOfFunctionSpaceByOPC}.

The eigenvalue of $\oH$ in zero-choreographic subspace
is doubly degenerated.
Because $\oH$ and $\oC$ commute,
if $\Psi_\lambda\ne0$ belongs to the eigenvalue $\oH'=\lambda$
then $\Psi_\lambda$ and
$\Phi_\lambda = (\oC-\oC^2)\Psi_\lambda/\sqrt{3}$
belong to the same eigenvalue.
Since $\Psi_\lambda$ satisfies $(1+\oC+\oC^2)\Psi_\lambda=0$,
$\braket{\Psi_\lambda|\Phi_\lambda}=0$
and $\braket{\Phi_\lambda|\Phi_\lambda}=\braket{\Psi_\lambda|\Psi_\lambda}\ne0$.
Actually,
\begin{equation}
\begin{split}
\braket{\Psi_\lambda|\Phi_\lambda}
&=\braket{\Psi_\lambda|\oC-\oC^2|\Psi_\lambda}/\sqrt{3}
=\big(\braket{\Psi_\lambda|\oC\Psi_\lambda}-\braket{\oC\Psi_\lambda|\Psi_\lambda}\big)/\sqrt{3}
=0,\\
\braket{\Phi_\lambda|\Phi_\lambda}
&=\braket{\Psi_\lambda|(\oC^2-\oC)(\oC-\oC^2)|\Psi_\lambda}/3
=\braket{\Psi_\lambda|1|\Psi_\lambda}.
\end{split}
\end{equation}
Therefore $\Psi_\lambda$ and $\Phi_\lambda$ are
orthogonal bases of two dimensional subspace for each $\lambda$.
This proves the statement of this paragraph.

\subsection{Time reversal symmetry}
Let $\oT$ be the time reversal operator $\oT(\delta q_k(t))=\delta q_k(-t)$,
$\tau$ be the operator for exchange of the second  and third row,
\begin{equation}
\tau=\left(\begin{array}{ccc}1 & 0 & 0 \\0 & 0 & 1 \\0 & 1 & 0\end{array}\right),
\end{equation}
and let $\oD=\tau\oT$.
Then the time reversal symmetry  in (\ref{timeReversalSymmetry}) 
for the figure-eight solution is
\begin{equation}
\oD \left(\begin{array}{c}
q_0(t) \\q_1(t) \\q_2(t)
\end{array}\right)
=-\left(\begin{array}{c}
q_0(t) \\q_1(t) \\q_2(t)
\end{array}\right).
\end{equation}

Since, $\oD^2=1$, this operator defines the orthogonal projection operator
\begin{equation}
\oP_{\oD}=\frac{1}{2}(1+\oD),\ \oP_\oD(\oP_\oD-1)=0.
\end{equation}
This operator decomposes the function space into two subspaces
with $\oP_\oD'=1$ (i.e. $\oD'=1$)
and with $\oP_\oD'=0$ ($\oD'=-1$).
The figure-eight solution belongs to the subspace with $\oD'=-1$.

\subsection{Time shift  symmetry in $T/2$}
Let $R^{1/2}$ be a time displacement operator
$R^{1/2}\delta q(t)=\delta q(t+T/2)$,
$\mu$ be the mirror operator with respect to $y$ axis
$\mu \tr(\delta q_k^x, \delta q_k^y)=\tr(-\delta q_k^x, \delta q_k^y)$
and $\oM=\mu R^{1/2}$.
Then the time shift symmetry in $T/2$ is
\begin{equation}
\oM \left(\begin{array}{c}
q_0(t) \\q_1(t) \\q_2(t)
\end{array}\right)
=\left(\begin{array}{c}
q_0(t) \\q_1(t) \\q_2(t)
\end{array}\right).
\end{equation}

Since $\oM$ satisfies $\oM^2=1$, the eigenvalues are $\pm 1$,
and the orthogonal projection operator $P_\oM$ is
\begin{equation}
P_\oM=\frac{1}{2}(1+\oM),\ P_\oM(P_\oM-1)=0.
\end{equation}
This operator decompose the function space
into two subspaces with $P_\oM'=1$ (i.e. $\oM'=1$)
and $P_\oM'=0$ ($\oM'=-1$).
The figure-eight solution belongs to the subspace with $P_\oM'=1$.

\begin{table}
\caption{Zero-center-of-mass subspace is decomposed into $2^3$ subspaces
by three projection operators $\oP_\oC$, $\oP_\oD$ and $\oP_\oM$.
Correspondence between the name of subspaces and the eigenvalues are shown.
The last $\pm$ sign correspond to two eigenvalue of $\oP_\oM'=1,\ 0$
respectively.
}
\label{nameOfSubspaces}
\centering
\begin{tabular}{c|cc}
&$\oP_\oC'=1$&$\oP_\oC'=0$\\
\hline
$\oP_\oD'=1$&choreographic cos $\pm$&zero-choreographic cos $\pm$\\
$\oP_\oD'=0$&choreographic sin $\pm$&zero-choreographic sin $\pm$
\end{tabular}
\end{table}

\subsection{Summary of  symmetries for the figure-eight solution}
In the section \ref{section:zeroAndTrivialMode}
and previous subsections,
we introduced operators $\sigma$, $\oC$, $\oD$, $\oM$
and corresponding projection operators $\oP_\sigma$,
$\oP_\oC$, $\oP_\oD$, $\oP_\oM$.
All these projection operators commute each other.
Therefore, these operators decompose the function space
into $2^4$ subspaces.
Actually, we don't divide the trivial subspace $\oP_\sigma'=1$.
The zero-centre-of-mass subspace $\oP_\sigma'=0$ will be divided into $2^3$ subspaces.
See table \ref{nameOfSubspaces}.
The figure-eight solution belongs to the subspace
$\oP_\sigma'=0$ and
$\oP_\oC'=1$ ($\oC'=1$) and
$\oP_\oD'=0$ ($\oD'=-1$) and
and $\oP_\oM'=1$ ($\oM'=1$).

The eigenspace of $\oH$ is also divided into these  subspaces,
because all these projection operators commute with $\oH$.

In the next section,
we decompose the function space by these four symmetries.

\section{Subspaces and base functions}
\label{sectionDecomposition}
Before  decompose the function space,
we count the degree of this space.
To make the degree finite, we introduce a cutoff
in Fourier series,
$1$, $\cos(\nu t)$, $\cos(2\nu t)$, $\cos(3\nu t),~\dots$,
$\cos(3N\nu t)$
and $\sin(\nu t), \sin(2\nu t), \sin(3\nu t)\dots, \sin(3N\nu t)$,
$\nu = 2\pi/T$
and $N=2^M$ with some integer $M$.
Then the degrees are $3$ for three bodies, $2$ for $x$ and $y$ components,
$6N+1$ for Fourier components.
So, total degree is $6(6N+1)\sim 2 \times 18N$.
We will decompose this space.
See figure \ref{blocksOfH}.
In this figure, trivial subspace is not decomposed.
The compliment, zero-centre-of-mass subspace, will be decomposed into $2^3$ subspaces.
\begin{figure}
{\small
\begin{equation*}
M_\oH=
\left(
\begin{array}{cccccccccccccccccc}
\cdot&&&&&&&&&&&&&&&&&\\
&\cdot&&&&&&&&&&&&&&&&\\
&&\cdot&&&&&&&&&&&&&&&\\
&&&\cdot&&&&&&&&&&&&&&\\
&&&&\cdot&&&&&&&&&&&&&\\
&&&&&\cdot&&&&&&&&&&&&\\
&&&&&&*&&&&&&&&&&&\\
&&&&&&&*&&&&&&&&&&\\
&&&&&&&&*&&&&&&&&&\\
&&&&&&&&&*&&&&&&&&\\
&&&&&&&&&&*&*&&&&&&\\
&&&&&&&&&&*&*&&&&&&\\
&&&&&&&&&&&&*&*&&&&\\
&&&&&&&&&&&&*&*&&&&\\
&&&&&&&&&&&&&&*&*&&\\
&&&&&&&&&&&&&&*&*&&\\
&&&&&&&&&&&&&&&&*&*\\
&&&&&&&&&&&&&&&&*&*
\end{array}\right)
\begin{array}{l}
\mbox{trivial}\\
\\
\\
\\
\\
\\
\mbox{choreographic cos }+\\
\phantom{\mbox{choreographic cos }}-\\
\mbox{choreographic sin }+\\
\phantom{\mbox{choreographic sin }}-\\
\mbox{zero-choreographic cos }+\\
\\
\mbox{zero-choreographic cos }-\\
\\
\mbox{zero-choreographic sin }+\\
\\
\mbox{zero-choreographic sin }-\\
\\
\end{array}
\end{equation*}
}
\caption{Decomposition of  matrix representation of the Hessian,
whose total size is $(2 \times 18N)^2$.
Here, each $\cdot$ stands $(2 \times N)^2$ diagonal matrix,
and $*$ stands $(2 \times N)^2$ matrix.
So, the size of each 4 choreographic blocks are $(2 \times N)^2$,
and zero-choreographic blocks are $(2 \times 2N)^2$.
All no marked elements are $0$.
}
\label{blocksOfH}
\end{figure}

Although we are considering the function space of $\mathbb{R}^6$,
it is convenient to consider the function space of $\mathbb{C}^6$
in the following two subsections.
This is because operator $\oO$ with $\oO^3=1$
has the eigenvalue only $\oO'=1$ in $\mathbb{R}$.
This makes our decomposition not straightforward.
It has three eigenvalues $\oO'=1, \omega, \omega^2$,
$\omega=(-1+i\sqrt{3})/2$ in $\mathbb{C}$  instead.
We will use $e^{-i 3N\nu t}, \dots, e^{-i \nu t}, 1, e^{i\nu t}, \dots, e^{i 3N\nu t}$
instead of $\sin$ and $\cos$ for the Fourier series.
We again have $6N+1$ Fourier components.

\subsection{Decomposition by $\oP_\sigma$}
Since $\sigma^3=1$,
the eigenvalue of $\sigma$ are $\sigma'=1, \omega, \omega^2$
with $\omega = (-1+i\sqrt{3})/2$.
Then, the basis for $\oP_\sigma=1$ are trivial,
\begin{equation}
\sigma'=1\leftrightarrow
\lleft(\begin{array}{c}1 \\1\\1 \\\end{array}\rright) f(t)
\end{equation}
with arbitrary periodic function $f(t)$.

While function for $\oP'=0$ are zero-centre-of-mass,
\begin{equation}
\sigma'=\omega \leftrightarrow
\lleft(\begin{array}{c}1 \\\omega\\\omega^2 \\\end{array}\rright) f(t),
\mbox{ or }
\sigma'=\omega^2 \leftrightarrow
\lleft(\begin{array}{c}1 \\\omega^2\\\omega \\\end{array}\rright) f(t).
\end{equation}
Therefore $1/3$ of whole basis are trivial
and $2/3$ are zero-centre-of-mass.
Namely, $2(6N+1)$ basis are trivial. We don't  decompose them more.

In the following sections, we will decompose 
$4(6N+1)$ basis by other projection operators.

\subsection{Decomposition by $\oP_\oC$}
Since $(R^{1/3})^3=1$, the eigenvalues of $R^{1/3}$
are $1, \omega$ and $\omega^2$.
The eigenfunctions are
$e^{i3k\nu t}, e^{i(3k+1)\nu t}$ and $e^{i(3k+2)\nu t}$
with $k \in \mathbb{Z}$
respectively.
More precisely,
\begin{equation}
\begin{split}
&(R^{1/3})'=1 \leftrightarrow
e^{i(-3N)\nu t}, e^{i(-3N+3)\nu t},\dots, e^{i(-3)\nu t}, 1, e^{i3\nu t},e^{i6\nu t},
\dots, e^{i(3N)\nu t},\\
&(R^{1/3})'=\omega \leftrightarrow
e^{i(-3N+1)\nu t}, e^{i(-3N+4)\nu t},\dots, e^{i(-2)\nu t},  e^{i\nu t},e^{i4\nu t},
\dots, e^{i(3N-2)\nu t},\\
&(R^{1/3})'=\omega^2 \leftrightarrow
e^{i(-3N+2)\nu t}, e^{i(-3N+5)\nu t},\dots, e^{-i\nu t}, e^{i2\nu t},e^{i5\nu t},
\dots, e^{i(3N-1)\nu t}.
\end{split}
\end{equation}
Each subspace has $2N+1$, $2N$ and $2N$ functions.

Then, choreographic basis belong to
$\oP_\oC'=1 \leftrightarrow \oC'=(\sigma^{-1})'(R^{1/3})'=1 \leftrightarrow
(\sigma', (R^{1/3})')=(\omega, \omega)$ or $(\omega^2, \omega^2)$.
Namely, the following $8N$ basis are choreographic,
\begin{equation}
\label{choreographicModes}
\lleft(\begin{array}{l}1 \\\omega\\\omega^2 \\\end{array}\rright)
e^{i(3k+1)\nu t}
\mbox{ and }
\lleft(\begin{array}{l}1 \\\omega^2\\\omega \\\end{array}\rright)
e^{i(3k+2)\nu t}.
\end{equation}
It should be stressed again that
\begin{equation}
\lleft(\begin{array}{l}1 \\\omega\\\omega^2 \\\end{array}\rright)
\end{equation}
represents two independent basis
\begin{equation}
\left(\begin{array}{l}1 \\0 \\\omega \\0 \\\omega^2 \\0\end{array}\right),\ 
\left(\begin{array}{l}0\\1\\0 \\\omega \\0 \\\omega^2\end{array}\right).
\end{equation}
So,
\begin{equation}
\lleft(\begin{array}{l}1 \\\omega\\\omega^2 \\\end{array}\rright)
e^{i(3k+1)\nu t},\ 
k=-N, -N+1,\dots, -1,0,1,\dots N-1
\end{equation}
stands for $2\times 2N$ base functions.

Zero-choreographic basis $\oP_\oC'=0$ are composed of
$\oC'=(\sigma^{-1})'(R^{1/3})'=\omega \mbox{ or }\omega^2$.
Namely,
$\oC'=(\sigma^{-1})'(R^{1/3})'=\omega
\leftrightarrow
(\sigma', (R^{1/3})')=(\omega, \omega^2) \mbox{ or } (\omega^2, 1)
\leftrightarrow$
\begin{equation}
\label{zeroChoreographicModesI}
\lleft(\begin{array}{l}1 \\\omega\\\omega^2 \\\end{array}\rright)
e^{i(3k+2)\nu t}
\mbox{ and }
\lleft(\begin{array}{l}1 \\\omega^2\\\omega \\\end{array}\rright)
e^{i3k\nu t},
\end{equation}
and
$\oC'=(\sigma^{-1})'(R^{1/3})'=\omega^2
\leftrightarrow
(\sigma', (R^{1/3})')=(\omega, 1), (\omega^2, \omega)
\leftrightarrow$
\begin{equation}
\label{zeroChoreographicModesII}
\lleft(\begin{array}{l}1 \\\omega\\\omega^2 \\\end{array}\rright)
e^{i3k\nu t}
\mbox{ and }
\lleft(\begin{array}{l}1 \\\omega^2\\\omega \\\end{array}\rright)
e^{i(3k+1)\nu t}.
\end{equation}
So, there are $16N+4$ zero-choreographic basis.

Thus,
total $4(6N+1)$ basis are split into $8N$ choreographic basis
and $16N+4$ zero-choreographic basis.

\subsection{Decomposition by $\oP_\oD$}
Now consider the action $\oD=\tau \oT$ to the above basis
(\ref{choreographicModes}),
(\ref{zeroChoreographicModesI}) and
(\ref{zeroChoreographicModesII}).
The operator $\tau$ exchange $\omega$ and $\omega^2$
and $\oT$ exchange $e^{ik\nu t}$ and $e^{-ik\nu t}$.
As the result, $\oD=\tau \oT$ exchange a base 
and its complex conjugate.
Therefore, the projection operator $\oP_\oD=(1+\oD)/2$
picks its real part,
while  $1-\oP_\oD=(1-\oD)/2$ picks its imaginary part.

Thus, the basis for
$(\oP_\sigma', \oP_\oC', \op_\oD')=(0,1,1)$ are
\begin{equation}
\lleft(\begin{array}{l}
\cos((3n+1)\nu t) \\\cos((3n+1)\nu t+2\pi/3) \\\cos((3n+1)\nu t-2\pi/3)\end{array}
\rright)
=
\lleft(\begin{array}{l}
\cos((3n+1)\nu t) \\\cos((3n+1)\nu (t+T/3)) \\\cos((3n+1)\nu (t-T/3)\end{array}
\rright)
\end{equation}
or
\begin{equation}
\lleft(\begin{array}{l}
\cos((3n+2)\nu t) \\\cos((3n+2)\nu t-2\pi/3) \\\cos((3n+2)\nu t+2\pi/3)\end{array}
\rright)
=
\lleft(\begin{array}{l}
\cos((3n+2)\nu t) \\\cos((3n+2)\nu (t+T/3)) \\\cos((3n+2)\nu (t-T/3)\end{array}
\rright).
\end{equation}
Explicitly,
\begin{equation}
\begin{split}
\lleft(\begin{array}{l}
\cos(\nu t) \\\cos(\nu (t+T/3)) \\\cos(\nu (t-T/3)\end{array}
\rright),&
\lleft(\begin{array}{l}
\cos(2\nu t) \\\cos(2\nu (t+T/3)) \\\cos(2\nu (t-T/3)\end{array}
\rright),\\
\lleft(\begin{array}{l}
\cos(4\nu t) \\\cos(4\nu (t+T/3)) \\\cos(4\nu (t-T/3)\end{array}
\rright),&
\lleft(\begin{array}{l}
\cos(5\nu t) \\\cos(5\nu (t+T/3)) \\\cos(5\nu (t-T/3)\end{array}
\rright),\\
&\vdots\\
\lleft(\begin{array}{l}
\cos((3N-2)\nu t) \\\cos((3N-2)\nu (t+T/3)) \\\cos((3N-2)\nu (t-T/3)\end{array}
\rright),&
\lleft(\begin{array}{l}
\cos((3N-1)\nu t) \\\cos((3N-1)\nu (t+T/3)) \\\cos((3N-1)\nu (t-T/3)\end{array}
\rright).
\end{split}
\end{equation}
Let $k_n=\{1,2,4,5,7,8,10,\dots,3N-2, 3N-1\}$, namely, 
the series of positive integers without multiple of 3.
This series is given by
\begin{equation}
k_n=\frac{1}{4}\Big( 6n-(-1)^n-3\Big),\ 
n=1,2,3,\dots,2N-1,2N.
\end{equation}
There are $4N$ basis for $(\oP_\sigma', \oP_\oC', \oD')=(0,1,1)$,
and $4N$ basis for $(\oP_\sigma', \oP_\oC',\oD')=(0,1,-1)$.
We write them
\begin{equation}
(\oP_\sigma', \oP_\oC', \oD')=(0,1,1)
\leftrightarrow
cc_n=\lleft(\begin{array}{l}
\cos(k_n\nu t) \\\cos(k_n\nu (t+T/3)) \\\cos(k_n\nu (t-T/3)\end{array}
\rright)
\end{equation}
and
\begin{equation}
(\oP_\sigma', \oP_\oC', \oD')=(0,1,-1)
\leftrightarrow
cs_n=\lleft(\begin{array}{l}
\sin(k_n\nu t) \\\sin(k_n\nu (t+T/3)) \\\sin(k_n\nu (t-T/3)\end{array}
\rright),
\end{equation}
for $n=1,2,3,\dots,2N-1,2N$.

For zero-choreographic subspace, $(\oC',\oD')=(0,1)$ basis are
\begin{equation}
\begin{split}
&\lleft(\begin{array}{r}1 \\-1/2 \\-1/2\end{array}\rright)\!\!,
\lleft(\begin{array}{l}\cos(\nu t) \\\cos(\nu t -2\pi/3) \\\cos(\nu t+2\pi/3)\end{array}\rright)\!\!,
\lleft(\begin{array}{l}\cos(2\nu t) \\\cos(2\nu t +2\pi/3) \\\cos(2\nu t-2\pi/3)\end{array}\rright)\!\!,
\lleft(\begin{array}{l}\cos(3\nu t) \\\cos(3\nu t -2\pi/3) \\\cos(2\nu t+2\pi/3)\end{array}\rright),\\
&\lleft(\begin{array}{l}\cos(3\nu t) \\\cos(3\nu t +2\pi/3) \\\cos(3\nu t-2\pi/3)\end{array}\rright)\!\!,
\lleft(\begin{array}{l}\cos(4\nu t) \\\cos(4\nu t -2\pi/3) \\\cos(4\nu t+2\pi/3)\end{array}\rright)\!\!,
\lleft(\begin{array}{l}\cos(5\nu t) \\\cos(5\nu t +2\pi/3) \\\cos(5\nu t-2\pi/3)\end{array}\rright)\!\!,
\lleft(\begin{array}{l}\cos(6\nu t) \\\cos(6\nu t -2\pi/3) \\\cos(6\nu t+2\pi/3)\end{array}\rright),\\
&\vdots\\
&\hdots,
\lleft(\begin{array}{l}\cos((3N-2)\nu t) \\\cos((3N-2)\nu t -2\pi/3) \\\cos((3N-2)\nu t+2\pi/3)\end{array}\rright)\!\!,
\lleft(\begin{array}{l}\cos((3N-1)\nu t) \\\cos((3N-1)\nu t +2\pi/3) \\\cos((3N-1)\nu t-2\pi/3)\end{array}\rright)\!\!,
\lleft(\begin{array}{l}\cos(3N\nu t) \\\cos(3N\nu t -2\pi/3) \\\cos(3N\nu t+2\pi/3)\end{array}\rright),\\
&\lleft(\begin{array}{l}\cos(3N\nu t) \\\cos(3N\nu t +2\pi/3) \\\cos(3N\nu t-2\pi/3)\end{array}\rright).
\end{split}
\end{equation}
Using series $\ell_n=\{0,1,2,3,3,4,5,6,6,7,8,9,9,10,11,\dots,3N-2,2N-1,3N,3N\}$
which is generated by
\begin{equation}
\begin{split}
\ell_n=\frac{1}{8}\Big( 6n+(-1)^n+(1-i)(-i)^n+(1+i)i^n-3\Big)\\
\mbox{for }n=1,2,3,\dots,4N+1,
\end{split}
\end{equation}
the basis are
\begin{equation}
\label{ZC}
(\oP_\sigma', \oP_\oC', \oD')=(0,0,1)
\leftrightarrow
zc_n=\lleft(\begin{array}{l}
	\cos(\ell_n \nu t) \\
	\cos(\ell_n \nu t -(-1)^n 2\pi/3) \\
	\cos(\ell_n \nu t +(-1)^n 2\pi/3)
	\end{array}\rright),
\end{equation}
for $n=1,2,3,\dots,4N+1$.
There are $4N+1$ basis.
Similarly,
\begin{equation}
\label{ZS}
(\oP_\sigma', \oP_\oC', \oD')=(0,0,-1)
\leftrightarrow
zs_n=(-1)^{n+1}\lleft(\begin{array}{l}
	\sin(\ell_n \nu t) \\
	\sin(\ell_n \nu t -(-1)^n 2\pi/3) \\
	\sin(\ell_n \nu t +(-1)^n 2\pi/3)
	\end{array}\rright),
\end{equation}
for $n=1,2,3,\dots,4N+1$.
There are again $4N+1$ basis.
The reason of the factor $(-1)^{n+1}$ will be shown soon.

We can easily check the relation
\begin{equation}
\label{rotationForZCandZS}
\oC \left(\begin{array}{c}zc_n \\zs_n\end{array}\right)
=\left(\begin{array}{rr}-1/2 & \sqrt{3}/2 \\-\sqrt{3}/2 & -1/2\end{array}\right)
\left(\begin{array}{c}zc_n \\zs_n\end{array}\right).
\end{equation}
Namely, the operator $\oC$ makes $2\pi/3$ rotation
in the subspace spanned by $zc_n$ and $zs_n$ for each $n$.
To prove this relation,
it is useful to note that
the behavior of $n, \ell_n$ and $\ell_n \nu T/3=2\pi\ell_n/3$ are 
summarized in the following table.
\begin{equation}
\begin{array}{ccccc}
n \bmod 4 & 1 & 2 & 3 & 0 \\
\ell_n \bmod 3 & 1 & 2 & 0 & 0 \\
2\pi\ell_n/3 \bmod 2\pi &2\pi/3 & -2\pi/3 & 0 & 0
\end{array}
\end{equation}
So, to prove the relation in (\ref{rotationForZCandZS}),
it is sufficient to prove the four cases $n=1,2,3,4$.
The factor in (\ref{ZS}) is chosen to satisfy (\ref{rotationForZCandZS}).

Now, consider a function $\Psi_+$ in $(\oP_oC', \oP_\oD')=(0,1)$
and $\Psi_-$ in $(\oP_oC', \oP_\oD')=(0,-1)$,
\begin{equation}
\Psi_{+}=\sum_n zc_n\ a_n,\ \Psi_{-}=\sum_n zs_n\ a_n
\end{equation}
with the same coefficients $a_n$.
Then, we have
\begin{equation}
\oC\left(\begin{array}{c}\Psi_{+} \\\Psi_{-}\end{array}\right)
=\left(\begin{array}{rr}-1/2 & \sqrt{3}/2 \\-\sqrt{3}/2 & -1/2\end{array}\right)\left(\begin{array}{c}\Psi_{+} \\\Psi_{-}\end{array}\right).
\end{equation}
So, if $\Psi_{+}$ belongs to $\oH'=\lambda$ then $\Psi_{-}$ belongs to the same $\lambda$,
because $\oC(\oH-\lambda)\Psi_{+}=0$ implies
\begin{equation}
0 
= (\oH-\lambda)\oC\Psi_{+}=(\oH-\lambda)\left(-\frac{1}{2}\Psi_{+} + \frac{\sqrt{3}}{2}\Psi_{-}\right)
=\frac{\sqrt{3}}{2}(\oH-\lambda)\Psi_{-}.
\end{equation}
Inverse is also true. This proves explicitly the double degeneracy of zero-choreographic subspace.

\subsection{Decomposition by $\oP_\oM$}
For integer $m$,
$\cos(m\nu (t+T/2))=\cos(m\nu t + m \pi)=(-1)^m \cos(m\nu t)$ and
$\sin(m\nu (t+T/2))=(-1)^m \sin(m\nu t)$.
So, for $x$ and $y$ components,
\begin{equation}
\oM'=1
\leftrightarrow
\left(\begin{array}{cc}
	\cos((2m+1)\nu t) & 0 \\
	0 & \cos(2m' \nu t)
\end{array}\right)\!\!,
\left(\begin{array}{cc}
	\sin((2m+1)\nu t) & 0 \\
	0 & \sin(2m' \nu t)
\end{array}\right)
\end{equation}
and
\begin{equation}
\oM'=-1
\leftrightarrow
\left(\begin{array}{cc}
	\cos(2m\nu t) & 0 \\
	0 & \cos((2m'+1)\nu t)
\end{array}\right)\!\!,
\left(\begin{array}{cc}
	\sin(2m\nu t) & 0 \\
	0 & \sin((2m'+1) \nu t)
\end{array}\right).
\end{equation}

Decompose $k_n$
\begin{equation}
k_n
=\{1,2,4,5,7,8,10,11,13,14,16,\dots,3N-2,3N-1\}
\end{equation}
into odd set and even set,
\begin{equation}
\begin{split}
k^o_n
&=\{1,5,7,11,13,17,\dots,3N-1\}\\
&=\frac{1}{2}\Big(6n+(-1)^n -3\Big),\ n=1,2,3,\dots, N
\end{split}
\end{equation}
and
\begin{equation}
\begin{split}
k^e_n
&=\{2,4,8,10,14,16,\dots,3N-2\}\\
&=\frac{1}{2}\Big(6n-(-1)^n -3\Big),\ n=1,2,3,\dots, N.
\end{split}
\end{equation}
Since $k^o_n \bmod 3=\{1,2,1,2,\dots\}$
and $k^e_n \bmod 3=\{2,1,2,1\dots\}$,
\begin{equation}
\label{sink2pidiv3}
\sin(k^o_n 2\pi/3)=(-1)^{n+1}\frac{\sqrt{3}}{2}
\mbox{ and }
\sin(k^e_n 2\pi/3)=(-1)^{n}\frac{\sqrt{3}}{2}.
\end{equation}
Similarly,
\begin{equation}
\ell_n
=\{0,1,2,3,3,4,5,6,6,7,\dots,3N-2,3N-1,3N,3N\}
\end{equation}
into odd set and even set,
\begin{equation}
\begin{split}
\ell^o_n
&=\{1,3,3,5,7,9,9,11,13,15,15,\dots,3N-1\}\\
&=\frac{1}{4}\Big(6n+(-1)^n -(1-i)(-i)^n-(1+i)i^n-3\Big),\ n=1,2,3,\dots, 2N
\end{split}
\end{equation}
and
\begin{equation*}
\begin{split}
\ell^e_n
&=\{0,2,4,6,6,8,10,12,12,\dots,3N-2,3N,3N\}\\
&=\frac{1}{4}\Big(6n+(-1)^n+(1-i)(-i)^n+(1+i)i^n-3\Big),\ n=1,2,3,\dots,2N+1.
\end{split}
\end{equation*}
\emph{To make the number of elements of $\ell^o_n$ and $\ell^e_n$ equal,
we just omit the last element of $\ell^e_n$. 
} So, we use
\begin{equation}
\begin{split}
\ell^e_n
&=\{0,2,4,6,6,8,10,12,12,\dots,3N-2,3N\}\\
&=\frac{1}{4}\Big(6n+(-1)^n+(1-i)(-i)^n+(1+i)i^n-3\Big),\ n=1,2,3,\dots,2N.
\end{split}
\end{equation}

The series with $\ell^o_n \nu t$ in (\ref{ZC}) is
\begin{equation}
\label{seriesNCOddn}
\begin{split}
\left(\begin{array}{l}\cos(\nu t) \\\cos(\nu t-2\pi/3) \\\cos(\nu t+2\pi/3)\end{array}\right),&\
\left(\begin{array}{l}\cos(3\nu t) \\\cos(3\nu t-2\pi/3) \\\cos(3\nu t+2\pi/3)\end{array}\right),\
\left(\begin{array}{l}\cos(3\nu t) \\\cos(3\nu t+2\pi/3) \\\cos(2\nu t-2\pi/3)\end{array}\right),\
\left(\begin{array}{l}\cos(5\nu t) \\\cos(5\nu t+2\pi/3) \\\cos(5\nu t-2\pi/3)\end{array}\right),\\
\left(\begin{array}{l}\cos(7\nu t) \\\cos(7\nu t-2\pi/3) \\\cos(7\nu t+2\pi/3)\end{array}\right),&\
\left(\begin{array}{l}\cos(9\nu t) \\\cos(9\nu t-2\pi/3) \\\cos(9\nu t+2\pi/3)\end{array}\right),\
\left(\begin{array}{l}\cos(9\nu t) \\\cos(9\nu t+2\pi/3) \\\cos(9\nu t-2\pi/3)\end{array}\right),\
\left(\begin{array}{l}\cos(11\nu t) \\\cos(11\nu t+2\pi/3) \\\cos(11\nu t-2\pi/3)\end{array}\right),\\
\dots\\
&=
\left(\begin{array}{l}
	\cos(\ell^o_n\nu t) \\
	\cos(\ell^o_n\nu t+(-1)^{\lfloor (n+1)/2 \rfloor} 2\pi/3) \\
	\cos(\ell^o_n\nu t-(-1)^{\lfloor (n+1)/2 \rfloor} 2\pi/3)
\end{array}\right)
\mbox{ for } n=1,2,3,\dots, 2N.
\end{split}
\end{equation}

Similarly, the series with $\ell^e_n \nu t$ in (\ref{ZC}) is
\begin{equation}
\label{seriesNCEvenn}
\begin{split}
\left(\begin{array}{r}1 \\-1/2 \\-1/2\end{array}\right),&\
\left(\begin{array}{l}\cos(2\nu t) \\\cos(2\nu t+2\pi/3) \\\cos(2\nu t-2\pi/3)\end{array}\right),\
\left(\begin{array}{l}\cos(4\nu t) \\\cos(4\nu t-2\pi/3) \\\cos(4\nu t+2\pi/3)\end{array}\right),\
\left(\begin{array}{l}\cos(6\nu t) \\\cos(6\nu t-2\pi/3) \\\cos(6\nu t+2\pi/3)\end{array}\right),\\
\left(\begin{array}{l}\cos(6\nu t) \\\cos(6\nu t+2\pi/3) \\\cos(6\nu t-2\pi/3)\end{array}\right),&\
\left(\begin{array}{l}\cos(8\nu t) \\\cos(8\nu t+2\pi/3) \\\cos(8\nu t-2\pi/3)\end{array}\right),\
\left(\begin{array}{l}\cos(10\nu t) \\\cos(10\nu t-2\pi/3) \\\cos(10\nu t+2\pi/3)\end{array}\right),\
\left(\begin{array}{l}\cos(12\nu t) \\\cos(12\nu t-2\pi/3) \\\cos(12\nu t+2\pi/3)\end{array}\right),\\
\dots\\
&=
\left(\begin{array}{l}
	\cos(\ell^e_n\nu t) \\
	\cos(\ell^e_n\nu t-(-1)^{\lfloor (n+1)/2 \rfloor} 2\pi/3) \\
	\cos(\ell^e_n\nu t+(-1)^{\lfloor (n+1)/2 \rfloor} 2\pi/3)
\end{array}\right)
\mbox{ for } n=1,2,3,\dots, 2N.
\end{split}
\end{equation}

\subsection{Summary for the base functions}
\subsubsection{Choreographic subspace}
Using $k^o_n$ and $k^e_n$,
\begin{equation}
(\oP_\sigma', \oP_\oC',\oD',\oM')=(0,1,1,1)\leftrightarrow
cc_n^+
=\left(\begin{array}{ll}
\cos(k^o_n \nu t) & 0 \\
0 & \cos(k^e_n \nu t) \\
\cos(k^o_n \nu (t+T/3)) & 0 \\
0 & \cos(k^e_n \nu (t+T/3)) \\
\cos(k^o_n \nu (t-T/3)) & 0 \\
0 & \cos(k^e_n \nu (t-T/3)) \\
\end{array}\right),
\end{equation}
and
\begin{equation}
(\oP_\sigma', \oP_\oC',\oD',\oM')=(0,1,1,-1)\leftrightarrow
cc_n^-
=\left(\begin{array}{ll}
\cos(k^e_n \nu t) & 0 \\
0 & \cos(k^o_n \nu t) \\
\cos(k^e_n \nu (t+T/3)) & 0 \\
0 & \cos(k^o_n \nu (t+T/3)) \\
\cos(k^e_n \nu (t-T/3)) & 0 \\
0 & \cos(k^o_n \nu (t-T/3)) \\
\end{array}\right),
\end{equation}
for $n=1,2,3,\dots,N$.

Similarly,
\begin{equation}
(\oP_\sigma', \oP_\oC',\oD',\oM')=(0,1,-1,1)\leftrightarrow
cs_n^+
=\left(\begin{array}{ll}
\sin(k^o_n \nu t) & 0 \\
0 & \sin(k^e_n \nu t) \\
\sin(k^o_n \nu (t+T/3)) & 0 \\
0 & \sin(k^e_n \nu (t+T/3)) \\
\sin(k^o_n \nu (t-T/3)) & 0 \\
0 & \sin(k^e_n \nu (t-T/3)) \\
\end{array}\right)
\end{equation}
and
\begin{equation}
(\oP_\sigma', \oP_\oC',\oD',\oM')=(0,1,-1,-1)\leftrightarrow
cs_n^-
=\left(\begin{array}{ll}
\sin(k^e_n \nu t) & 0 \\
0 & \sin(k^o_n \nu t) \\
\sin(k^e_n \nu (t+T/3)) & 0 \\
0 & \sin(k^o_n \nu (t+T/3)) \\
\sin(k^e_n \nu (t-T/3)) & 0 \\
0 & \sin(k^o_n \nu (t-T/3)) \\
\end{array}\right)
\end{equation}
for $n=1,2,3,\dots,N$.

It may useful to define a function
\begin{equation}
\Phi_c(f, k^x, k^y, n)
=\left(\begin{array}{ll}
f(k^x_n\nu t) & 0 \\
0 & f(k^x_n\nu t)  \\
f(k^x_n\nu (t+T/3)) & 0 \\
0 & f(k^x_n\nu (t+T/3))\\
f(k^x_n\nu (t-T/3)) & 0 \\
0 & f(k^x_n\nu (t-T/3)t)
\end{array}\right).
\end{equation}
Then,
\begin{equation}
\begin{array}{ll}
cc^+_n=\Phi_c(\cos, k^o, k^e,n),&cc^-_n=\Phi_c(\cos, k^e, k^o,n),\\
cs^+_n=\Phi_c(\sin, k^o, k^e,n),&cc^-_n=\Phi_c(\sin, k^e, k^o,n)
\end{array}
\end{equation}

\subsubsection{Zero-choreographic subspace}
\begin{equation}
\begin{split}
&(\oP_\sigma', \oP_\oC', \oD',\oM')=(0,0,1,1)\leftrightarrow\\
&zc^+_n=\left(\begin{array}{ll}
		\cos(\ell^o_n\nu t) & 0 \\
		0 & \cos(\ell^e_n\nu t) \\
		\cos(\ell^o_n\nu t+(-1)^{\lfloor (n+1)/2\rfloor} 2\pi/3) & 0 \\
		0 & \cos(\ell^e_n\nu t-(-1)^{\lfloor (n+1)/2\rfloor} 2\pi/3) \\
		\cos(\ell^o_n\nu t-(-1)^{\lfloor (n+1)/2\rfloor} 2\pi/3) & 0 \\
		0 & \cos(\ell^e_n\nu t+(-1)^{\lfloor (n+1)/2\rfloor} 2\pi/3)
		\end{array}\right),
\end{split}
\end{equation}
and
\begin{equation}
\begin{split}
&(\oP_\sigma', \oP_\oC', \oD',\oM')=(0,0,1,-1)\leftrightarrow\\
&zc^-_n=\left(\begin{array}{ll}
		\cos(\ell^e_n\nu t) & 0 \\
		0 & \cos(\ell^o_n\nu t) \\
		\cos(\ell^e_n\nu t-(-1)^{\lfloor (n+1)/2\rfloor} 2\pi/3) & 0 \\
		0 & \cos(\ell^o_n\nu t+(-1)^{\lfloor (n+1)/2\rfloor} 2\pi/3) \\
		\cos(\ell^e_n\nu t+(-1)^{\lfloor (n+1)/2\rfloor} 2\pi/3) & 0 \\
		0 & \cos(\ell^o_n\nu t-(-1)^{\lfloor (n+1)/2\rfloor} 2\pi/3)
		\end{array}\right)
\end{split}
\end{equation}
for $n=1,2,3,\dots, 2N$.
Similarly,
\begin{equation}
\begin{split}
&(\oP_\sigma', \oP_\oC', \oD',\oM')=(0,0,-1,1)\leftrightarrow\\
&zs^+_n=(-1)^{n+1}\left(\begin{array}{ll}
		\sin(\ell^o_n\nu t) & 0 \\
		0 & \sin(\ell^e_n\nu t) \\
		\sin(\ell^o_n\nu t+(-1)^{\lfloor (n+1)/2\rfloor} 2\pi/3) & 0 \\
		0 & \sin(\ell^e_n\nu t-(-1)^{\lfloor (n+1)/2\rfloor} 2\pi/3) \\
		\sin(\ell^o_n\nu t-(-1)^{\lfloor (n+1)/2\rfloor} 2\pi/3) & 0 \\
		0 & \sin(\ell^e_n\nu t+(-1)^{\lfloor (n+1)/2\rfloor} 2\pi/3)
		\end{array}\right),
\end{split}
\end{equation}
and
\begin{equation}
\begin{split}
&(\oP_\sigma', \oP_\oC', \oD',\oM')=(0,0,-1,-1)\leftrightarrow\\
&zs^-_n=(-1)^{n+1}\left(\begin{array}{ll}
		\sin(\ell^e_n\nu t) & 0 \\
		0 & \sin(\ell^o_n\nu t) \\
		\sin(\ell^e_n\nu t-(-1)^{\lfloor (n+1)/2\rfloor} 2\pi/3) & 0 \\
		0 & \sin(\ell^o_n\nu t+(-1)^{\lfloor (n+1)/2\rfloor} 2\pi/3) \\
		\cos(\ell^e_n\nu t+(-1)^{\lfloor (n+1)/2\rfloor} 2\pi/3) & 0 \\
		0 & \sin(\ell^o_n\nu t-(-1)^{\lfloor (n+1)/2\rfloor} 2\pi/3).
		\end{array}\right)
\end{split}
\end{equation}
for $n=1,2,3,\dots, 2N$.

It may useful to define a function
\begin{equation}
\Psi_z(f, \ell^x, \ell^y, \epsilon, n)
=\left(\begin{array}{ll}
f(\ell^x_n \nu t) & 0 \\
0 & f(\ell^y_n \nu t) \\
f(\ell^x_n \nu t + \epsilon (-1)^{\lfloor (n+1)/2\rfloor} 2\pi/3) & 0\\
0 & f(\ell^y_n \nu t -\epsilon (-1)^{\lfloor (n+1)/2\rfloor} 2\pi/3)\\
f(\ell^x_n \nu t - \epsilon (-1)^{\lfloor (n+1)/2\rfloor} 2\pi/3) & 0\\
0 & f(\ell^y_n \nu t +\epsilon (-1)^{\lfloor (n+1)/2\rfloor} 2\pi/3)
\end{array}\right).
\end{equation}
Then, the base functions are
\begin{equation}
\begin{array}{ll}
zc^+_n=\Psi(\cos, \ell^o, \ell^e, 1,n),\ 
&zc^-_n=\Psi(\cos, \ell^e, \ell^o, -1,n),\\
zs^+_n=(-1)^{n+1}\Psi(\sin, \ell^o, \ell^e, 1,n),\ 
&zs^-_n=(-1)^{n+1}\Psi(\sin, \ell^e, \ell^o, -1,n).
\end{array}
\end{equation}

\section{Matrix elements of Hessian}
\subsection{Definition of matrix elements}
Let $\phi_n$ be $6 \times 2$ matrix that describes bases 
and $a_n$ be $2$ column vector that describes degrees of freedom.
For,
\begin{equation}
\phi_n = \left(\begin{array}{cc}
	\delta q_{0n}^x & 0 \\
	0 & \delta q_{0n}^y \\
	\delta q_{1n}^x & 0 \\
	0 & \delta q_{1n}^y \\
	\delta q_{2n}^x & 0 \\
	0 & \delta q_{2n}^y
	\end{array}
\right)
\ 
\mbox{ and }
a_n=\left(\begin{array}{c}a_{n}^x \\a_{n}^y\end{array}\right),
\end{equation}
the product $\phi_n a_n$ represents the vector
\begin{equation}
\phi_n a_n
=\left(\begin{array}{c}
	\delta q_{0n}^x a_{n}^x\\
	\delta q_{0n}^y a_{n}^y\\
	\delta q_{1n}^x a_{n}^x\\
	\delta q_{1n}^y a_{n}^y\\
	\delta q_{2n}^x a_{n}^x\\
	\delta q_{2n}^y a_{n}^y
	\end{array}\right).
\end{equation}
Note that we have only two degrees of freedom $a_{n}^x$ and $a_{n}^y$ for 
given $n$.

Now, consider the eigenvalue problem
\begin{equation}
\oH \sum_{n=1,2,3,\dots} \phi_n a_n = \lambda \sum_{n=1,2,3,\dots} \phi_n a_n.
\end{equation}
Substituting the expression
\begin{equation}
\sum_{n=1,2,3,\dots} \phi_n a_n
=(\phi_1, \phi_2, \phi_3, \dots)\left(\begin{array}{c}a_1 \\ a_2 \\ a_3 \\\vdots\end{array}\right)
\end{equation}
into the eigenvalue problem,
and multiply ${}^t\!(\phi_1, \phi_2, \phi_3, \dots)$ from the left
and taking the inner product,
we get the eigenvalue problem in a matrix representation
\begin{equation}
M_\oH \left(\begin{array}{c}a_1 \\ a_2 \\ a_3 \\\vdots\end{array}\right)
	=\lambda \left(\begin{array}{c}a_1 \\ a_2 \\ a_3 \\\vdots\end{array}\right).
\end{equation}
Where $M_\oH$ is the matrix representation of the Hessian,
\begin{equation}
M_\oH=
\left(\begin{array}{cccc}
	\braket{\phi_1|\oH|\phi_1} & \braket{\phi_1|\oH|\phi_2} & \braket{\phi_1|\oH|\phi_3} & \dots\\
	\braket{\phi_2|\oH|\phi_1} & \braket{\phi_2|\oH|\phi_2} & \braket{\phi_2|\oH|\phi_3} & \dots\\
	\braket{\phi_3|\oH|\phi_1} & \braket{\phi_3|\oH|\phi_2} & \braket{\phi_3|\oH|\phi_3} & \dots\\
	\vdots & \vdots & \vdots & \ddots
	\end{array}\right).
\end{equation}
The inner product is defined by (\ref{defOfInnerProduct}).

\subsection{Preparations}
Since we are considering the Hessian
at a choreography $q_0(t)=q(t)$, $q_1(t)=q(t+T/3)$ and $q_2(t)=q(t-T/3)$,
the function $u_{ij}$ satisfy
$u_{20}(t)=u_{12}(t+T/3)$ and $u_{01}(t)=u_{12}(t+T/3)$.
Let $u_{12}(t)=u(t)$, then $u_{20}(t)=R^{1/3}u(t)R^{-1/3}$ and
$u_{01}(t)=R^{-1/3}u(t)R^{1/3}$.
Then, ${}^t\!\Delta \oU \Delta$ in (\ref{defOfHessian}) is
\begin{equation}
{}^t\!\Delta \oU \Delta
={}^t\!\Delta
	\left(\begin{array}{ccc}
		1 & 0 & 0 \\
		0 & R^{1/3} & 0 \\
		0 & 0 & R^{-1/3}
	\end{array}\right)
	\left(\begin{array}{ccc}
		u & 0 & 0 \\
		0 & u & 0 \\
		0 & 0 & u
	\end{array}\right)
	\left(\begin{array}{ccc}
		1 & 0 & 0 \\
		0 & R^{-1/3} & 0 \\
		0 & 0 & R^{1/3}
	\end{array}\right)
	\Delta.
\end{equation}
In the following subsections, we use the notation
\begin{equation}
u=\left(\begin{array}{cc}u^{xx} & u^{xy} \\u^{yx} & u^{yy}\end{array}\right).
\end{equation}
We will use the lower case letters $a, b, \dots$ to express $x$ or $y$.

\subsection{Elements for choreographic subspace}
For the matrix elements for choreographic subspace,
consider
\begin{equation}
\begin{split}
&\left(\begin{array}{ccc}
		1 & 0 & 0 \\
		0 & R^{-1/3} & 0 \\
		0 & 0 & R^{1/3}
	\end{array}\right)
	\Delta
	\left(\begin{array}{l}
		\cos(k\nu t) \\\cos(k\nu(t+T/3)) \\\cos(k\nu(t-T/3))\end{array}
	\right)\\
&=\left(\begin{array}{ccc}
		1 & 0 & 0 \\
		0 & R^{-1/3} & 0 \\
		0 & 0 & R^{1/3}
	\end{array}\right)
	\left(\begin{array}{l}
		\cos(k\nu(t+T/3))-\cos(k\nu(t-T/3)) \\
		\cos(k\nu(t-T/3))-\cos(k\nu t) \\
		\cos(k\nu t)-\cos(k\nu(t+T/3))
	\end{array}\right)\\
&=\big(\cos(k\nu(t+T/3))-\cos(k\nu(t-T/3))\big)\left(\begin{array}{c}1 \\1 \\1\end{array}\right)\\
&=-2 \sin(2\pi k/3)\sin(k\nu t){}^t\!(1,1,1).
\end{split}
\end{equation}
Similarly,
\begin{equation}
\begin{split}
&\left(\begin{array}{ccc}
		1 & 0 & 0 \\
		0 & R^{-1/3} & 0 \\
		0 & 0 & R^{1/3}
	\end{array}\right)
	\Delta
	\left(\begin{array}{l}
		\sin(k\nu t) \\\sin(k\nu(t+T/3)) \\\sin(k\nu(t-T/3))\end{array}
	\right)\\
&=\big(\sin(k\nu(t+T/3))-\sin(k\nu(t-T/3))\big) {}^t\!(1,1,1)\\
&=2 \sin(2\pi k/3)\cos(k\nu t) {}^t\!(1,1,1).
\end{split}
\end{equation}
Therefore,
$ab$ component of 
$\braket{\Phi_c(\cos,k^x,k^y,m|\tr\Delta\oU\Delta|\Phi(k^x,k^y,n)}$
is 
\begin{equation}
\begin{split}
&\braket{\Phi_c(\cos,k^x,k^y,m|\tr\Delta\oU\Delta|\Phi(k^x,k^y,n)}^{ab}\\
&=\sin(2\pi k^a_m/3)\sin(2\pi k^b_n/3)
\times\frac{8}{T}\int_0^T dt\ u^{ab}(t)
	\sin(k^a_m\nu t)\sin(k^b_n\nu t)
\end{split}
\end{equation}
The relation (\ref{sink2pidiv3}) yields
\begin{equation}
\begin{split}
&\sin(2\pi k^o_m/3)\sin(2\pi k^o_n/3)
=\sin(2\pi k^e_m/3)\sin(2\pi k^e_n/3)
=(-1)^{m+n}\frac{3}{4},\\
&\sin(2\pi k^o_m/3)\sin(2\pi k^e_n/3)
=-(-1)^{m+n}\frac{3}{4}.
\end{split}
\end{equation}
So, we finally get the matrix elements
\begin{equation}
\label{uforCCPlus}
\begin{split}
&\braket{cc_m^+|{}^t\!\Delta \oU \Delta |cc_n^+}\\
&=(-1)^{m+n}\frac{6}{T}\int_0^T dt
\left(\begin{array}{rr}
	u^{xx}\sin(k^o_m\nu t)\sin(k^o_n\nu t) & -u^{xy}\sin(k^o_m\nu t)\sin(k^e_n\nu t) \\
	-u^{xy}\sin(k^e_m\nu t)\sin(k^o_n\nu t) & u^{yy}\sin(k^e_m\nu t)\sin(k^e_n\nu t)
\end{array}\right),
\end{split}
\end{equation}
\begin{equation}
\label{uforCCMinus}
\begin{split}
&\braket{cc_m^-|{}^t\!\Delta \oU \Delta |cc_n^-}\\
&=(-1)^{m+n}\frac{6}{T}\int_0^T dt
\left(\begin{array}{rr}
	u^{xx}\sin(k^e_m\nu t)\sin(k^e_n\nu t) & -u^{xy}\sin(k^e_m\nu t)\sin(k^o_n\nu t) \\
	-u^{xy}\sin(k^o_m\nu t)\sin(k^e_n\nu t) & u^{yy}\sin(k^o_m\nu t)\sin(k^o_n\nu t)
\end{array}\right),
\end{split}
\end{equation}
\begin{equation}
\label{uforCSPlus}
\begin{split}
&\braket{cs_m^+|{}^t\!\Delta \oU \Delta |cs_n^+}\\
&=(-1)^{m+n}\frac{6}{T}\int_0^T dt
\left(\begin{array}{rr}
	u^{xx}\cos(k^o_m\nu t)\cos(k^o_n\nu t) & -u^{xy}\cos(k^o_m\nu t)\cos(k^e_n\nu t) \\
	-u^{xy}\cos(k^e_m\nu t)\cos(k^o_n\nu t) & u^{yy}\cos(k^e_m\nu t)\cos(k^e_n\nu t)
\end{array}\right),
\end{split}
\end{equation}
and
\begin{equation}
\label{uforCSMinus}
\begin{split}
&\braket{cs_m^-|{}^t\!\Delta \oU \Delta |cs_n^-}\\
&=(-1)^{m+n}\frac{6}{T}\int_0^T dt
\left(\begin{array}{rr}
	u^{xx}\cos(k^e_m\nu t)\cos(k^e_n\nu t) & -u^{xy}\cos(k^e_m\nu t)\cos(k^o_n\nu t) \\
	-u^{xy}\cos(k^o_m\nu t)\cos(k^e_n\nu t) & u^{yy}\cos(k^o_m\nu t)\cos(k^o_n\nu t)
\end{array}\right).
\end{split}
\end{equation}
The differences between the above elements are 
only the place of $k^o$ and $k^e$.

\subsection{Elements for zero-choreographic subspace}
To calculate the $ab$ element of
$\braket{zc^+_m|\tr\Delta\oU\Delta|zc^+}$,
let $\epsilon^x=(-1)^{\lfloor (n+1)/2\rfloor}$
and $\epsilon^y=-(-1)^{\lfloor (n+1)/2\rfloor}$.
Then
\begin{equation}
\begin{split}
&\left(\begin{array}{ccc}1 & 0 & 0 \\0 & R^{-1/3} & 0 \\0 & 0 & R^{1/3}\end{array}\right)
\!\Delta 
\left(\begin{array}{l}
	\cos(\ell^a_n \nu t) \\
	\cos(\ell^a_n \nu t + 2\pi \epsilon^a_n/3) \\
	\cos(\ell^a_n \nu t - 2\pi \epsilon^a_n/3)
\end{array}\right)\\
&=\sqrt{3}\epsilon^a_n
	\left(\begin{array}{l}
	\sin(\ell^a_n \nu t)\\
	\sin(\ell^a_n \nu (t-T/3) - \pi \epsilon^a_n/3)\\
	\sin(\ell^a_n \nu (t+T/3)+\pi \epsilon^a_n/3)
\end{array}\right).
\end{split}
\end{equation}

Therefore,
\begin{equation}
\begin{split}
&\braket{zc^+_m|\tr\Delta \oU \Delta|zc^+_n}^{ab}\\
=&
\epsilon^a_m \epsilon^b_n 
\frac{1}{T}\int_0^T dt\ u^{ab}(t)\Bigg(
	\cos\Big((\ell^a_m-\ell^b_n)\nu t\big) 
	- \cos\big((\ell^a_m+\ell^b_n)\nu t\big)\\
	&+\cos\Big((\ell^a_m-\ell^b_n)\nu (t-T/3)-(\epsilon^a_m-\epsilon^b_n)\pi/3\Big)
	-\cos\Big((\ell^a_m+\ell^b_n)\nu (t-T/3)-(\epsilon^a_m+\epsilon^b_n)\pi/3\Big)\\
	&+\cos\Big((\ell^a_m-\ell^b_n)\nu (t+T/3)+(\epsilon^a_m-\epsilon^b_n)\pi/3\Big)
	-\cos\Big((\ell^a_m+\ell^b_n)\nu (t+T/3)+(\epsilon^a_m+\epsilon^b_n)\pi/3\Big)
	\Bigg)\\
=&\epsilon^a_m \epsilon^b_n \frac{3}{T}\int_0^T dt\ 
u(t)L(\ell^a, \ell^b, \epsilon^a, \epsilon^b, m, n).
\end{split}
\end{equation}
The term $L(\ell^a, \ell^b, \epsilon^a, \epsilon^b, m, n)$ turns out to be
\begin{equation}
L(\ell^a, \ell^b, \epsilon^a, \epsilon^b, m, n)
=(-1)^{m+n}\cos\Big( ((-1)^m \ell^a_m -(-1)^n\ell^b_n)\nu t).
\end{equation}
Using
$(-1)^{\lfloor (n+1)/2 \rfloor}(-1)^n
=1,-1,-1,1,1,-1,-1,1,1,\dots=(-1)^{\lfloor n/2 \rfloor}$.

Finally, we get
\begin{equation}
\begin{split}
&\braket{nc^+_m|{}^t\!\Delta \oU \Delta| nc^+_n} 
=(-1)^{\lfloor m/2 \rfloor + \lfloor n/2 \rfloor} \frac{3}{T}\\
&\times\int_0^T \!dt
\left(\begin{array}{rr}
u^{xx}(t)\cos\Big(((-1)^m\ell^o_m-(-1)^n\ell^o_n)\nu t\Big) & 
	-u^{xy}(t)\cos\Big(((-1)^m\ell^o_m-(-1)^n\ell^e_n)\nu t\Big) \\
-u^{yx}(t)\cos\Big(((-1)^m\ell^e_m-(-1)^n\ell^o_n)\nu t\Big) &
u^{yy}(t)\cos\Big(((-1)^m\ell^e_m-(-1)^n\ell^e_n)\nu t\Big)\end{array}
\right).
\end{split}
\end{equation}
Exchanging $\ell^o \leftrightarrow \ell^e$,
we get $\braket{nc^-_m|{}^t\!\Delta \oU \Delta| nc^-_n}$,
\begin{equation}
\begin{split}
&\braket{nc^-_m|{}^t\!\Delta \oU \Delta| nc^-_n} 
=(-1)^{\lfloor m/2 \rfloor + \lfloor n/2 \rfloor} \frac{3}{T}\\
&\times\int_0^T \!dt
\left(\begin{array}{rr}
u^{xx}(t)\cos\Big(((-1)^m\ell^e_m-(-1)^n\ell^e_n)\nu t\Big) & 
	-u^{xy}(t)\cos\Big(((-1)^m\ell^e_m-(-1)^n\ell^o_n)\nu t\Big) \\
-u^{yx}(t)\cos\Big(((-1)^m\ell^o_m-(-1)^n\ell^e_n)\nu t\Big) &
u^{yy}(t)\cos\Big(((-1)^m\ell^o_m-(-1)^n\ell^o_n)\nu t\Big)\end{array}
\right).
\end{split}
\end{equation}

Similarly,
\begin{equation}
\begin{split}
&\left(\begin{array}{ccc}1 & 0 & 0 \\0 & R^{-1/3} & 0 \\0 & 0 & R^{1/3}\end{array}\right)
\!\Delta 
\left(\begin{array}{l}
	\sin(\ell^a_n \nu t) \\
	\sin(\ell^a_n \nu t + 2\pi \epsilon^a_n/3) \\
	\sin(\ell^a_n \nu t - 2\pi \epsilon^a_n/3)
\end{array}\right)\\
&=\sqrt{3}\epsilon^a_n
	\left(\begin{array}{l}
	\cos(\ell^a_n \nu t)\\
	-\cos(\ell^a_n \nu (t-T/3) - \pi \epsilon^a_n/3)\\
	-\cos(\ell^a_n \nu (t+T/3)+\pi \epsilon^a_n/3)
\end{array}\right).
\end{split}
\end{equation}

Therefore,
\begin{equation}
\begin{split}
&\braket{zs^+_m|\tr\Delta \oU \Delta|zs^+_n}^{ab}\\
=&
(-1)^{m+n}
\epsilon^a_m \epsilon^b_n 
\frac{1}{T}\int_0^T dt\ u^{ab}(t)\Bigg(
	\cos\Big((\ell^a_m-\ell^b_n)\nu t\big) 
	+ \cos\big((\ell^a_m+\ell^b_n)\nu t\big)\\
	&+\cos\Big((\ell^a_m-\ell^b_n)\nu (t-T/3)-(\epsilon^a_m-\epsilon^b_n)\pi/3\Big)
	+\cos\Big((\ell^a_m+\ell^b_n)\nu (t-T/3)-(\epsilon^a_m+\epsilon^b_n)\pi/3\Big)\\
	&+\cos\Big((\ell^a_m-\ell^b_n)\nu (t+T/3)+(\epsilon^a_m-\epsilon^b_n)\pi/3\Big)
	+\cos\Big((\ell^a_m+\ell^b_n)\nu (t+T/3)+(\epsilon^a_m+\epsilon^b_n)\pi/3\Big)
	\Bigg)\\
=&(-1)^{m+n}\epsilon^a_m \epsilon^b_n \frac{3}{T}\int_0^T dt\ 
u(t)L'(\ell^a, \ell^b, \epsilon^a, \epsilon^b, m, n).
\end{split}
\end{equation}
The term $L'(\ell^a, \ell^b, \epsilon^a, \epsilon^b, m, n)$ turns out to be
\begin{equation}
L'(\ell^a, \ell^b, \epsilon^a, \epsilon^b, m, n)
=\cos\Big( ((-1)^m \ell^a_m -(-1)^n\ell^b_n)\nu t).
\end{equation}
Thus, we directly get
\begin{equation}
\begin{split}
\braket{zs^+_m|\tr\Delta\oU\Delta|zs^+_n}
&=\braket{zc^+_m|\tr\Delta\oU\Delta|zc^+_n},\\
\braket{zs^-_m|\tr\Delta\oU\Delta|zs^-_n}
&=\braket{zc^-_m|\tr\Delta\oU\Delta|zc^-_n}.
\end{split}
\end{equation}
This is a direct proof of the double degeneracy of $\oH$
in $\oP_\oC'=0$ subspace.

\subsection{Summary for the matrix elements}
Let $\tilde{u}^{ab}(k)$ be
\begin{equation}
\left(\begin{array}{cc}
\tilde{u}^{xx}(k) & \tilde{u}^{xy}(k) \\
\tilde{u}^{yx}(k) & \tilde{u}^{yy}(k)
\end{array}\right)
=\frac{1}{T}\int_0^T \!\!dt\ 
\left(\begin{array}{cc}
u^{xx}(t) & u^{xy}(t) \\
u^{yx}(t) & u^{yy}(t)
\end{array}\right)
\cos(k\nu t),
\end{equation}
$k=0,1,2,\dots,6N-1$, $N=2^M$.

\subsubsection{Choreographic subspace}
Let $C(k^x, k^y,\epsilon, m,n)$ be
\begin{equation}
C(k^x, k^y,\epsilon, m,n)
=3(-1)^{m+n}\left(\begin{array}{rr}
\tilde{u}^{xx}(k^x+\epsilon k^x) & -\tilde{u}^{xy}(k^x+\epsilon k^y)\\
-\tilde{u}^{yx}(k^y+\epsilon k^x) & \tilde{u}^{yy}(k^y+\epsilon k^y)
\end{array}\right).
\end{equation}

Subspace: $(\oP_\oC',\oP_\oD',\oP_\oM')=(0,1,1)$,\\
name: Choreographic cos$+$,\\
symbol: $cc^+$,\\
range: $m,n=1,2,3,\dots, N$,\\
matrix element:
\begin{equation}
\braket{cc^+_m|\tr\Delta\oU\Delta|nc^+_n}
=C(k^o,k^e,-1,m,n)-C(k^o,k^e,1,m,n).
\end{equation}

Subspace: $(\oP_\oC',\oP_\oD',\oP_\oM')=(0,1,-1)$,\\
name: Choreographic cos$-$,\\
symbol: $cc^-$,\\
range: $m,n=1,2,3,\dots, N$,\\
matrix element:
\begin{equation}
\braket{cc^-_m|\tr\Delta\oU\Delta|nc^-_n}
=C(k^e,k^o,-1,m,n)-C(k^e,k^o,1,m,n).
\end{equation}

Subspace: $(\oP_\oC',\oP_\oD',\oP_\oM')=(0,-1,1)$,\\
name: Choreographic sin$+$,\\
symbol: $cs^+$,\\
range: $m,n=1,2,3,\dots, N$,\\
matrix element:
\begin{equation}
\braket{cs^+_m|\tr\Delta\oU\Delta|ns^+_n}
=C(k^o,k^e,-1,m,n)+C(k^o,k^e,1,m,n).
\end{equation}

Subspace: $(\oP_\oC',\oP_\oD',\oP_\oM')=(0,-1,-1)$,\\
name: Choreographic sin$-$,\\
symbol: $cs^-$,\\
range: $m,n=1,2,3,\dots, N$,\\
matrix element:
\begin{equation}
\braket{cs^-_m|\tr\Delta\oU\Delta|ns^-_n}
=C(k^e,k^o,-1,m,n)+C(k^e,k^o,1,m,n).
\end{equation}

\subsubsection{Zero-choreographic subspace}
Let $Z(\ell^x, \ell^y,m,n)$ be
\begin{equation}
\begin{split}
Z(\ell^x, \ell^y,m,n)
=&3(-1)^{\lfloor m/2\rfloor+\lfloor n/2\rfloor}\\
&\times
\left(\begin{array}{rr}
\tilde{u}^{xx}((-1)^m \ell^x_m-(-1)^n \ell^x_n)&
-\tilde{u}^{xy}((-1)^m \ell^x_m-(-1)^n \ell^y_n) \\
-\tilde{u}^{yx}((-1)^m \ell^y_m-(-1)^n \ell^x_n)&
\tilde{u}^{yy}((-1)^m \ell^y_m-(-1)^n \ell^y_n)
\end{array}\right).
\end{split}
\end{equation}

Subspace: $(\oP_\oC',\oP_\oD',\oP_\oM')=(0,\pm 1,1)$,\\
name: Zero-choreographic cos$+$ and sin$+$,\\
symbol: $zc^+$ and $zs^+$,\\
range: $m,n=1,2,3,\dots, 2N$,\\
matrix element:
\begin{equation}
\braket{zc^+_m|\tr\Delta\oU\Delta|zc^+_n}
=\braket{zs^+_m|\tr\Delta\oU\Delta|zs^+_n}
=Z(\ell^o,\ell^e,m,n).
\end{equation}

Subspace: $(\oP_\oC',\oP_\oD',\oP_\oM')=(0,\pm 1,-1)$,\\
name: Zero-choreographic cos$-$ and sin$-$,\\
symbol: $zc^-$ and $zs^-$,\\
range: $m,n=1,2,3,\dots, 2N$,\\
matrix element:
\begin{equation}
\braket{zc^-_m|\tr\Delta\oU\Delta|zc^-_n}
=\braket{zs^-_m|\tr\Delta\oU\Delta|zs^-_n}
=Z(\ell^e,\ell^o,m,n).
\end{equation}
.

\section{Discussions}
\subsection{Scaling and coupling dependence of eigenvalues}
Consider the eigenvalue problem $\oH\Psi=\lambda \Psi$, namely,
\begin{equation}
\left( -\frac{d^2}{dt^2}+{}^t\!\Delta \oU \Delta \right)\Psi = \lambda \Psi
\end{equation}
for extended Newton potential in (\ref{homogeneousPotential}).

The eigenvalues scales $\lambda \to \lambda/\mu^2$ for
the scaling  of $t \to \mu t$ and $q \to \mu^{2/(2+\alpha)}q$.
Therefore, it is useful to consider the scale invariant eigenvalue $\tilde{\lambda}$
defined by
\begin{equation}
\tilde{\lambda}=\frac{\lambda}{4\pi^2/T^2}.
\end{equation}
Here, the factor $4\pi^2$ is for later convenience.

Consider a system under a homogeneous potential with coupling constant $g^2>0$, 
described by
\begin{equation}
\label{generalizedLagrangian}
L=1/2\sum (dq_\ell/dt)^2 + g^2 \sum V_\alpha(|q_i-q_j|).
\end{equation}
The Lagrangian in (\ref{Lagrangian}) has $g^2=1$.
Then the corresponding eigenvalue problem is
\begin{equation}
\left( -d^2/dt^2+g^2\ {}^t\!\Delta \oU \Delta \right)\Psi = \lambda \Psi.
\end{equation}
The eigenvalue for the same period $T$ is the same for $g^2=1$,
because a scale transformation
$t \to t$ and $q \to g^{2/(2+\alpha)} q$ makes $g^2 \to 1$.
Therefore, the scale invariant eigenvalue $\tilde{\lambda}$ is
also invariant for changing coupling constant $g^2>0$.

\subsection{Numerical calculation}
In this subsection, we describe
a detail to calculate the eigenvalues and eigenfunctions 
of the Hessian for choreographic sine plus subspace 
to which the figure-eight solution belongs.

We introduced a cut-off of the Fourier series
$1, \cos(\nu t), \cos(2\nu t),\dots, \cos(3N\nu t)$
and 
$\sin(\nu t), \sin(2\nu t),\dots, \sin(3N\nu t)$
with sufficient large $N=2^M$.
Then, the largest number of $k^o_n$ and $k^e_n$ are
\begin{equation}
\label{maxForKoandKe}
k^o_n=\{1,5,7,\dots, k^o_{N}=3N-1\},\ 
k^e_n=\{2,4,8,\dots, k^e_{N}=3N-2\}.
\end{equation}
Using the equations
\begin{equation}
\begin{split}
&2\sin(k\nu t)\sin(k'\nu t)=-\cos((k+k')\nu t)+\cos((k-k')\nu t),\\
&2\cos(k\nu t)\cos(k'\nu t)=\cos((k+k')\nu t)+\cos((k-k')\nu t).
\end{split}
\end{equation}
the matrix elements (\ref{uforCCPlus}), (\ref{uforCCMinus}),
(\ref{uforCSPlus}) and (\ref{uforCSMinus})
are expressed by the Fourier integral
\begin{equation}
\label{Fourier}
\tilde{u}(k)=\frac{1}{T}\int_0^T u(t) \cos(k\nu t).
\end{equation}
The largest $k$ to calculate the matrix elements in (\ref{uforCSPlus})
for (\ref{maxForKoandKe})
is $k_\text{max}=2k^o_{N}=6N-2 < 6N-1$.
We calculate the Fourier integral by  Fast Fourier Transformation,
\begin{equation}
\tilde{u}(k) = 
\operatorname{Re}
\left(\frac{1}{N}\sum_{0 \le s \le 6N-1} u\left(\frac{Ts}{N}\right)
				\exp(2\pi i k s/N)
\right).
\end{equation}
By Nyquist-Shannon-Someya sampling theorem,
we have to take the sampling number larger than $2 k_\text{max}$
to calculate $\tilde{u}(k)$ for  $0 \le k \le k_\text{max}$ correctly.

Then, the elements (\ref{uforCSPlus}) are
\begin{equation}
\begin{split}
&\braket{cs_m^+|{}^t\!\Delta \oU \Delta |cs_n^+}\\
&=(-1)^{m+n}3
\left(\begin{array}{rr}
\tilde{u}_{xx}(k^o_m+k^o_n)+\tilde{u}_{xx}(k^o_m-k^o_n) & 
	-\tilde{u}_{xy}(k^o_m+k^e_n)-\tilde{u}_{xy}(k^o_m-k^e_n) \\
-\tilde{u}_{xy}(k^e_m+k^o_n)-\tilde{u}_{xy}(k^e_m-k^o_n) & 
	+\tilde{u}_{yy}(k^e_m+k^e_n)+\tilde{u}_{yy}(k^e_m-k^e_n)
\end{array}\right).
\end{split}
\end{equation}
Then, make the Hessian, 
$H_{mn}=\braket{cs_m^+|-d^2/dt^2|cs_n^+}+\braket{cs_m^+|{}^t\!\Delta U \Delta |cs_n^+}$,
and solve the eigenvalue problem $H \Phi = \lambda \Phi$.

Inversely, once we get the eigenvalue $\lambda$ and the eigenfunction
\begin{equation}
\Phi=\left(\begin{array}{c}a_1 \\a_2 \\a_3 \\\vdots \\a_{N}\end{array}\right),
\end{equation}
the eigenfunction $\Psi$ is given by
\begin{equation}
\Psi 
= \sum_{1 \le n \le N}cs^+_n a_n
=\left(\begin{array}{c} \delta q(t)\\\delta q(t+T/3) \\\delta q(t-T/3)\end{array}\right),
\end{equation}
where
\begin{equation}
\delta q(t)=\sum_{1 \le n \le N}
\left(\begin{array}{cc}\sin(k^o_n \nu t) & 0 \\0 & \sin(k^e_n \nu t)\end{array}\right)
		\left(\begin{array}{c}a_{n}^x \\a_{n}^y\end{array}\right)
=\sum_{1 \le n \le N}
	\left(\begin{array}{c}
	\sin(k^o_n \nu t)a_{n}^x \\
	\sin(k^e_n \nu t)a_{n}^y.
\end{array}\right).
\end{equation}
Each component of the right hand side is a Fourier series.
Using a series $b_k$, defined by
\begin{equation}
b_{k_n}=a_n \mbox{ for } n=1, 2, 3, \dots, N, \mbox{ all other } b_k=0,
\end{equation}
the series is given by
\begin{equation}
\begin{split}
\delta q(t)
&=\sum_{1 \le n \le N}\sin(k_n\nu t) a_n
=\sum_{0 \le k \le 3N-1} \sin(k\nu t) b_k\\
&=-\operatorname{Im} \left(\sum_{0 \le k \le 3N-1}  b_k \exp(-2\pi i kt/T)\right).
\end{split}
\end{equation}
We can use Inverse Fast Fourier Transform to calculate the right hand side numerically.
Namely, for discrete time $t=Ts/(3N)$, and $s=0,1,2, \dots, 3N-1$,
\begin{equation}
\delta q\left( \frac{Ts}{3N} \right)
=-\operatorname{Im} \left(\sum_{0 \le k \le 3N-1}  b_k \exp(-2\pi i ks/(3N))\right).
\end{equation}

\subsection{Application to the figure-eight solution}
Lower 5 scale invariant eigenvalues $\tilde{\lambda}=\lambda/(4\pi^2/T^2)$
for Newton potential ($\alpha=1$)
are listed in the  equation (\ref{eigenValuesForFigureEight}).
\begin{equation}
\label{eigenValuesForFigureEight}
\begin{array}{llllll}
cc+& 1.14719\times10^{-10}& 4.04112& 16.5764& 26.117& 50.469\\
cc-& 1.93391& 7.26091& 19.289& 27.3167& 50.6671\\
cs+& 2.32495& 6.14928& 18.5694& 27.9174& 50.9476\\
cs-& -6.81259\times10^{-10}& 3.54142& 16.4395& 25.7645& 50.272\\
zc+\mbox{ and }zs+& 0.00174304& 0.946257& 5.02738& 8.91162& 12.5666\\
zc-\mbox{ and }zs-& -0.0744809& 1.24605& 4.92153& 9.14293& 11.6371
\end{array}
\end{equation}
Tiny eigenvalues of order $10^{-10}$ in $cc+$ and $cs-$ are
zero eigenvalues that correspond to the conservation of
energy and angular momentum respectively.

This calculation is done with $N=2^9$.
(For definition of $N$, see section \ref{sectionDecomposition}
and figure \ref{blocksOfH}).

\subsection{A question for eigenvalue problem by differential equations}
The eigenvalue problem $\oH\Psi = \lambda \Psi$ can be solved 
by directly solving the differential equation 
(\ref{defOfHessian}) and (\ref{eigenValueProblem}) with periodic boundary condition.
Mathematica provides  ``NDEigensystem'' function for this purpose.
This function gives us an easy method to calculate the eigenvalues and
eigenfunctions in a moderate precision.
It is good idea to use this function if you need a moderate precision.
Actually, we used this function twice.
The first time is to get rough estimates of eigenvalues and eigenfunctions at an early stage of our calculations.
The second time is to check our program developed here at the last stage.

In these calculations,
we eliminated the variable $\delta q_2$ by $\delta q_2=-(\delta q_0+\delta q_1)$
to eliminate the trivial subspace.
And we impose periodic boundary condition $\delta q_k(t+T)=\delta q_k(t)$.
Equivalently,
we restricted the function space to zero-centre-of-mass 
and periodic. It works fine.

\emph{We have one question}.
Can we  restrict the function space more and more in this direct calculation
by imposing suitable boundary conditions?
For example,
we can restrict the function space to choreographic subspace
by imposing the boundary conditions
\begin{equation}
\delta q_1(t)=\delta q_0(t+T/3),\ \delta q_2(t)=\delta q_1(t+T/3),\ \delta q_0(t)=\delta q_2(t+T/3).
\end{equation}
The question is how to realize this boundary condition in numerical calculations,
for example in ``NDEigensystem''?
And how to realize  boundary conditions for choreographic cos $+$ in calculations?
A strong motivation for this note
is to classify eigenvalues and eigenvectors in suitable subspace defined by the symmetry.
If we can directly solve the eigenvalue equation in each subspace
with realizing suitable boundary conditions,
this will provide a concise and simple method.
It will be a future investigation.

\section*{Acknowledgements}
We would like to express many thanks to M.~Shibayama
for his work \cite{Shibayama}
that describes a general method to calculate the eigenvalue of the Hessian.
This work was partly supported by JSPS Grant-in-Aid for Scientific Research 
17K05146 (HF) and 17K05588(HO).


\begin{thebibliography}{99}
\bibitem{Moore}C.~Moore 1993,
		``Braids in Classical Dynamics'',
		Phys. Rev. Lett. 70 3675
\bibitem{CM}A.~Chenciner and R.~Montgomery 2000, 
		``A remarkable periodic solution of the three-body problem in the case of equal masses'',
		Ann. of Math. (2) 1523, 881--901

\bibitem{Sbano}L.~Sbano 2004,
		``Symmetric solutions in molecular potentials'',
		Symmetry and Perturbation Theory,
	Proceedings of the International Conference SPT 2004,
	World Scientific


\bibitem{Shibayama}M.~Shibayama 2010, 
	``Numerical calculation of the second variation for the choreographic solution'', in Japanese,
	Computations and Calculations in Celestial Mechanics,
	Proceedings of Symposium on Celestial Mechanics and N-body Dynamics (2010),
	Eds. M.~Saito, M.~Shibayama and M.~Sekiguchi.


\bibitem{Suvakov}M.~\v{S}uvakov and V.~Dmitra\v{s}inovi\'{c} 2013, 
		``Three Classes of Newtonian Three-Body Planar Periodic Orbits'',
		Phys. Rev. Lett. 110 114301
\bibitem{SuvakovShibayama}M.~\v{S}uvakov and M.~Shibayama 2016,
		``Three topologically nontrivial choreographic motions of three bodies'',
		Celest. Mech. Dyn. Astron. 124 155-162

\bibitem{LJ1}H.~Fukuda, T.~Fujiwara, H.~Ozaki 2017,
	``Figure-eight choreographies of the equal mass three-body problem 
	with Lennard-Jones-type potentials'',
	J. Phys. A: Math. Theor.
\end{thebibliography}
\end{document}